\documentclass[12pt,preprint]{aastex}

\slugcomment{The Astrophysical Journal, in press}
\shorttitle{Anisotropic Locations of Satellite Galaxies}
\shortauthors{Agustsson \& Brainerd}

\begin{document}

\title{Anisotropic Locations of Satellite Galaxies: Clues to
the Orientations of Galaxies within their Dark Matter Halos}

\author{Ingolfur Agustsson \& Tereasa G. Brainerd}
\affil{Boston University, Institute for Astrophysical Research, 725
Commonwealth Ave., Boston, MA 02215}
\email{ingolfur@bu.edu, brainerd@bu.edu}

\begin{abstract}
We investigate the locations of the satellites of relatively isolated
host galaxies in the Sloan Digital Sky Survey and the Millennium
Run simulation.  Provided we use two distinct prescriptions to embed luminous
galaxies within the simulated dark matter halos (ellipticals share the
shapes of their halos, while disks have angular momenta that are aligned
with the net angular momenta of their halos),
we find a fair agreement between observation and theory.
Averaged over scales $r_p \le 500$~kpc,
the satellites of red, high--mass hosts with low
star formation rates are found preferentially near the major axes of
their hosts.  In contrast, the satellites of blue, low--mass hosts with
low star formation rates show little to no anisotropy when averaged over the
same scale.  The difference between the locations of the satellites of
red and blue hosts cannot be explained by the effects of interlopers
in the data.
Instead, it is caused primarily by marked differences in the
dependence of the mean satellite
location, $\left< \phi \right>$, on the projected distance at which the
satellites are found.  We also find that
the locations of red, high--mass satellites with low star formation rates
show considerably more anisotropy than do the locations of
blue, low--mass satellites with high star formation rates.  There are two
contributors to this result.  First, the blue satellites
have only recently arrived within their hosts' halos, while the red
satellites arrived in the far distant past.  Second, the sample
of blue satellites is heavily contaminated by interlopers, which suppresses
the measured anisotropy compared to the intrinsic anisotropy.
\end{abstract}

\keywords{
dark matter --- galaxies: dwarf --- galaxies: fundamental
parameters --- galaxies: halos --- galaxies: structure}

\section{Introduction}
The locations of satellite galaxies, measured with respect to the symmetry
axes of their hosts, may hold important clues to the formation
of large galaxies.  This is especially true for Cold Dark Matter 
(CDM) models in which the dark matter halos of galaxies are
mildly flattened, and galaxy formation and mass accretion occur within
filaments.  Some early studies of the locations of satellite galaxies
suggested that satellites had a preference for being located near the
minor axes of their hosts (e.g., Holmberg 1969; Zaritsky et al.\ 1997), an observation
that is sometimes known as the ``Holmberg effect''.    
Valtonen et al.\ (1978) found exactly the opposite effect, and concluded that
compact satellites tended to be aligned with the major axes of their hosts.
Other early studies suggested
that any tendency for satellite galaxies to be found in preferred locations
was at best rather weak, and perhaps non--existent (e.g., Hawley \& Peebles 1975;
Sharp et al.\ 1979; MacGillivray et al.\ 1982).  
All of these early studies were based on relatively small samples of between
$\sim 10$ and $\sim 200$ satellite galaxies and as
modern, extensive redshift surveys have
become available, the observed number of host--satellite systems has
increased enormously.  Based upon these large surveys it
now appears that, when averaged over all host--satellite pairs, the 
satellites of relatively isolated host galaxies
have a tendency to be found near the
major axes of their hosts (see, e.g., Brainerd 2005).
There is, however, increasing evidence that the locations of the satellites  
depend upon host type (e.g., red vs.\ blue), as well as satellite type.

In an analysis of the locations of the satellites of relatively isolated
host galaxies in the Two Degree Field Galaxy Redshift Survey (2dFGRS; Colless
et al.\ 2001, 2003), Sales \& Lambas (2004) found a tendency for the
satellites of early--type hosts to be located near the major axes
of the hosts, while the satellites of late--type hosts were consistent
with being distributed isotropically (see the erratum by Sales \& Lambas
2009). In addition, they found a tendency for
the locations of
satellites with low star formation rates to show a greater degree of 
anisotropy than satellites with high star formation rates. 
Azzaro et al.\ (2007, hereafter APPZ) concluded that, as a whole, 
the satellites of relatively isolated host galaxies in the Sloan
Digital Sky Survey (SDSS; Fukugita et al.\ 1996; Hogg et al.\ 2001; Smith
et al.\ 2002; Strauss et al.\ 2002; York et al.\ 2000) were found preferentially
near the major axes of their hosts.  Further, APPZ found that the degree of
anisotropy was greatest for the red satellites of red host galaxies, while
the locations of the satellites of blue host galaxies were consistent with
an isotropic distribution.  Similar results were found by Siverd et al.\ (2009)
in a more recent analysis of the SDSS, where they showed that the satellites
of red, centrally--concentrated hosts are found preferentially close to the
major axes of the hosts, and the effect is strongest for red, centrally--concentrated
satellites.  In a study of extremely isolated SDSS host galaxies, Bailin
et al.\ (2008) found that the satellites of spheroidal host galaxies were
located preferentially close to the major axes of the hosts, while the satellites
of blue disk hosts were distributed isotropically.

The dependence of
satellite location on the color of the host has also been observed
within group environments by Yang et al.\ 
(2006, hereafter Yang06),
who found that the satellites of red central galaxies in
the SDSS had
a strong tendency to be aligned with the major axes of the central galaxies, while
the satellites of blue central galaxies were distributed isotropically about the
central galaxies.  Further, Yang06 found that the red satellites of
red central galaxies were distributed much 
more anisotropically than were the blue
satellites of red central galaxies, and the degree of anisotropy in the 
satellite locations increased only weakly
with the mass of the surrounding halo.

Here we further investigate the anisotropic distribution
of satellite galaxies around relatively isolated hosts,  focussing
on the dependence of the anisotropy on various physical parameters 
of the hosts and the satellites (e.g., rest--frame color, specific star formation
rate and stellar mass).   We also investigate the effects of ``interlopers''
(i.e., false satellites) on the locations of the satellites, as well
as the dependence
of satellite location on projected distance from the host.
The locations of satellites 
in the observed Universe are computed using SDSS galaxies, and these are
compared to the locations of satellites in the
$\Lambda$CDM Millennium Run simulation.  
Our work here is similar in spirit to that of Kang et al.\ (2007; hereafter Kang07), who
used a simulation that combined N-body calculations with
semi-analytic galaxy formation
to compare the locations of satellite
galaxies in a $\Lambda$CDM universe to the results obtained by Yang06
for SDSS satellites.  Our work differs from that of Kang07
in a number of ways, however.  First, we focus on the satellites of
relatively isolated host galaxies whereas
Yang06 and Kang07 focus 
primarily on group systems.
Second, in our work we use the stellar masses
of the host and satellite galaxies when exploring the dependence of the 
satellite locations on mass.  In contrast, Yang06 and Kang07
use a group luminosity function to assign masses to the dark matter halos
that surround their groups.
Third, we divide our theoretical
galaxies into two broad classes, elliptical and non--elliptical, and we
use different prescriptions to assign shape parameters to the luminous
portions of these galaxies.  Kang07, however, did not divide their
theoretical galaxies into different classes and they 
used identical prescriptions
to assign shape parameters to the luminous portions of all of their galaxies.

We note that Sales et al.\ (2007) have also investigated the locations of
satellite galaxies of relatively isolated host galaxies in the Millennium
Run.  Their approach, however, was rather different than our own.  Sales
et al.\ (2007) use the full information of the simulation (in particular, 
3D distances) to select their hosts and satellites, while we focus on samples
that are selected using the same selection criteria that 
are used to select hosts and satellites from large redshift surveys.  Having
full 3D information, Sales et al.\ (2007) selected all satellites 
with $M_r < -17$ that were 
found within the virial radii of their hosts and computed the locations of 
the satellites.  The result was preference for the satellites to populate
a plane that is perpendicular to the angular momentum axis of the host's
halo (i.e., the reverse of the Holmberg effect).
 
The outline of the paper is as follows. In \S2 we describe the
SDSS data, the Millennium Run simulation, and the way in which we
define images for the luminous host galaxies in the Millennium
Run.  In \S3 we discuss the
selection criteria for finding hosts and satellites, and we highlight some
of the properties of the host and satellite galaxies in the Millennium Run.
In \S4 we compute the locations of the satellite galaxies and we compare
the results obtained with SDSS galaxies to those obtained with the Millennium
Run galaxies.  We summarize our results and compare them
to previous, similar studies in
\S5, and we present our conclusions in \S6.
Throughout we adopt cosmological parameters 
$H_0 = 73$~km~sec$^{-1}$~Mpc$^{-1}$, $\Omega_{m0} = 0.25$, and
$\Omega_{\Lambda 0} =
0.75$.

\section{Observational and Theoretical Data Sets}

Our goal in this paper is to compute the locations of the
satellites of relatively isolated host galaxies for: (i) observed
galaxies in our Universe and (ii) theoretical galaxies in a $\Lambda$CDM
universe.  Below we outline the details of the observational and theoretical
data sets that are used in our analysis.

\subsection{Observed Galaxies: SDSS}

The SDSS is a large imaging and spectroscopic survey that has
mapped roughly one quarter of the sky. The spectroscopic portion of the
SDSS is complete to a reddening--corrected 
Petrosian magnitude of $r=17.77$ (see, e.g., Strauss et al. 2002).  Our primary
observational data set consists of the
seventh data release of the SDSS (DR7; Abazajian et al.\ 2009), including
all of the photometric and spectroscopic information for objects with high
quality redshifts (zconf $>$ 0.9) that have galaxy--type spectra
(specClass = 2), $r\le 17.77$, and redshifts in the range $0.01 \le z \le 0.15$.

We use the de--reddened Petrosian
$ugriz$ magnitudes (e.g., petroMag\_r-extinction\_r), and we 
select the
position angles, semi--minor axes, and semi--major
axes of our galaxies from the Petrosian $r$--band data.  In addition,
the IDL code by Blanton et al. (2003; v4\_1\_4) was used
to K-correct the SDSS galaxy colors to the present epoch (i.e., $z=0$).
Further, in some of the analyzes below we will supplement the data provided
directly by the SDSS with stellar mass estimates and
star formation rates.
Stellar masses are available for the 
vast majority of the galaxies in the DR7, but at the moment star formation
rates are only available for galaxies in the fourth SDSS data release (DR4;
Adelman-McCarthy et al.\ 2006).  Therefore, our galaxy sample will necessarily
be restricted when we look at the dependence of satellite location on
star formation rate. The 
stellar masses and star formation rates for
the SDSS galaxies are publicly available at {\tt http://www.mpa-garching.mpg.de/SDSS/}.
Stellar masses in these catalogs were computed using the philosophy of
Kauffmann et
al.\ (2003) and Salim et al.\ (2007). Star formation rates
were computed using various emission
lines in the SDSS spectra as described in Brinchmann et al.\
(2004). Throughout our analysis we use the specific star formation
rate (SSFR) of the SDSS galaxies, which is
defined to be the ratio of the star formation rate (in $M_\odot$~yr$^{-1}$) to
the stellar mass (in solar units), and we use the average values of
the likelihood distributions of the total SSFR obtained by
Brinchmann et al.\ (2004).

\subsection{Theoretical Galaxies: Millennium Run Simulation}

The Millennium Run simulation\footnote{http://www.mpa-garching.mpg.de/millennium}
(MRS) follows the
growth of cosmic structure in a $\Lambda$CDM ``concordance'' cosmology
($H_0 = 73$~km~sec$^{-1}$~Mpc$^{-1}$,$\Omega_{m0} + \Omega_{b0} = 0.25$,
$\Omega_{b0} = 0.04$, $\Omega_{\Lambda 0} = 0.75$, $n=1$,
$\sigma_8 = 0.9$).  The simulation was
completed by the Virgo Consortium in summer 2004 using the Max
Planck Society's supercomputer center in Garching, Germany, and
is described in Springel et al.\ (2005).
The simulation follows the evolution of the dark matter
distribution from $z = 127$ to $z = 0$ using $N = 2160^3 \simeq
10^{10}$ particles of mass $m_p = 8.6 \times 10^8 h^{-1} M_{\sun}$.
The simulation volume is a cubical box with periodic 
boundary conditions and a comoving side length of $L = 500 h^{-1}$~Mpc.
A TreePM method is used to evaluate the 
gravitational force law, and a softening length of $5
h^{-1}$~kpc is used.  The simulation thus achieves a truly impressive dynamic
range of $10^5$ in length. 
Since one of our goals is to construct an
accurate catalog of simulated host galaxies and their satellites, 
it is important for us to use a high--resolution
simulation that follows the fate of satellite galaxies accurately
as they orbit within the halo of the 
central host galaxy. The combination of high spatial and mass resolution
therefore makes the MRS
ideal for our purposes.

The stored output of the MRS allows semi--analytic models
of galaxy formation to be implemented by collecting
the detailed assembly histories of all resolved halos and
subhalos, then simulating the formation and
evolution of galaxies within these structures for a variety of
assumptions about the physics that is involved. The data on the
halo, subhalo, and galaxy populations which have been produced by such
efforts can be used to address a wide range of questions
about galaxy and structure evolution
(e.g., Croton et al.\ 2006). As part of the activities of the
German Astrophysical Virtual Observatory, detailed
information about the halos, subhalos, and galaxies have been
publicly released for two independent models of galaxy
formation (Lemson et al.\ 2006). 

In order to compare to the SDSS,
we need to analyze the MRS in the same way 
in which one would analyze a combined imaging and redshift survey of the
observed Universe.   To do this, we make use of the MRS all--sky mock galaxy redshift
catalog\footnote{http://www.g-vo.org/Millennium/Help?page=databases/mpamocks/blaizot2006\_\_allsky}
that was constructed by Blaizot et al.\ (2005) using the Mock
Map Facility (MoMaF).   The MRS mock redshift survey is intended to 
mimic the SDSS, having a nearly identical redshift distribution and very similar
color distributions for the galaxies.   The mock redshift survey incorporates
the semi--analytic galaxy formation model of De Lucia \& Blaizot (2007) for the
MRS galaxies. Therefore, galaxy fluxes in all of the SDSS bandpasses, as well as star formation
rates, stellar masses,
and $B$--band bulge--to--disk ratios, are available for the MRS galaxies.

In order to make the most direct comparison to
the SDSS, we need to include the galaxy images 
that one would have in a real observational survey.  That is, our goal is
to determine the locations of satellite galaxies, measured with respect to
the major axes of the images of their luminous host galaxies.
There are, however, no actual {\it images} of the simulated galaxies, and
we must therefore {\it define} images for the MRS host galaxies.  
As an aid to defining the image shapes, the bulge--to--disk
ratios from the semi--analytic galaxy formation model may be used
to assign rough intrinsic morphologies to the MRS hosts.
Following De Lucia et al. (2006) we therefore use the $B$--band
bulge--to--disk ratios to classify MRS host galaxies with $\Delta M(B)
< 0.4$ as ellipticals, where $\Delta M(B) =
M(B)_{\rm bulge}-M(B)_{\rm total}$.  Similarly, we classify MRS host galaxies with 
$\Delta M(B) \geq 0.4$ as ``non--ellipticals''.
We also note that visual inspection
of the images of the SDSS host galaxies has revealed these objects to 
be ``regular'' systems (i.e.,
ellipticals, lenticulars, or spirals).  Therefore, it is reasonable to assume
that the non--elliptical MRS hosts
are disk systems with
significant net angular momentum, and we will treat all non--elliptical MRS
hosts as though they were disk galaxies below.

Following Heavens et al. (2000) we assume that 
elliptical MRS host galaxies share the shapes of their dark matter halos.  During a
collaborative visit to the Max Planck Institute for Astrophysics (MPA), we were 
fortunate to be granted access to the particle data files that resulted from
the MRS.  The enormous size of the particle files precludes them from being made 
publicly-available; thus, at present,
it is only possible to work with the files on site at MPA.   During the 
visit to MPA the particles within the virial radii ($r_{200}$)  of the
elliptical MRS host galaxies were identified, and these particles 
were then used to compute 
equivalent ellipsoids of inertia for the elliptical hosts.  A total of 98\%
of the elliptical MRS hosts contain more than 1000 particles within their
virial radii, so the equivalent ellipsoids of inertia are well--determined.
The major axes of projections of these equivalent ellipsoids of
inertia onto the sky then
define the orientations of the major axes of the elliptical MRS host galaxies.  

In the case of the non--elliptical MRS hosts, it is natural to assume that the
net angular momentum of the disk will be perpendicular to the disk.
In addition, recent numerical simulations have indicated 
that the angular momenta of disk galaxies
and their dark matter halos are reasonably well--aligned (e.g., Libeskind
et al.\ 2007). Furthermore, the disk angular momentum vectors show a
tendency to be aligned with the minor axis of the surrounding mass
with a mean misalignment of $\sim 25^{\circ}$ (Bailin \&
Steinmetz 2005).  We, therefore,
computed the angular momentum vectors of the halos of the non--elliptical
MRS hosts using all particles contained within the virial radii.  These were
then used to place thin disks within the halos, oriented such
that the disks are perpendicular to the net angular momenta of the halos.
The major axes of the projections of these thin disks onto the sky then
define the orientations of the major axes of the non--elliptical MRS hosts.  
We note that
the angular momentum vectors of the host halos are well--determined, and 62\% of the
hosts contain more than 1000 particles that were used to compute the angular
momentum.  

\section{Host--Satellite Catalogs}

Although the MRS contains full 6--dimensional phase space information (i.e., positions
and velocities) for all of the galaxies, such is of course not the case for the observed 
Universe.  That is, since there is no direct distance information for the vast majority
of the galaxies in the SDSS, we are forced to select host
galaxies and their satellites using proximity criteria in redshift space, rather
than real space.  Again, in order to compare the simulation results as directly
as possible to the results from the SDSS, we select host and satellite galaxies
in the MRS in the same way that they are selected in the SDSS.  Below we discuss
our selection criteria and the resulting catalogs.

\subsection{Host--Satellite Selection Criteria}

Hosts and satellites are selected by requiring the hosts to be
relatively isolated.  In addition, hosts and satellites must be nearby 
one another in terms of projected separation on the sky, $r_p$,
and radial velocity difference, $|dv|$. Throughout we use the Sample 1 selection
criteria from Brainerd (2005). Specifically, hosts must be 2.5
times more luminous than any other galaxy that falls within $r_p
\leq 700$~kpc and $|dv| \leq 1000$~km~sec$^{-1}$. Satellites must
be at least 6.25 times less luminous than their host, and they must be
located within $r_p \leq 500$~kpc and $|dv| \leq 500$~kpc.
In order to eliminate a small number of systems that
pass the above tests but which are, in reality, more likely to be
representative of cluster environments instead of relatively isolated
host--satellite systems, we impose two further restrictions: (1)
the sum total of the luminosities of the satellites of a given
host must be less than the luminosity of the host, and (2) the
observed total number of satellites of a given host must not
exceed 9. Our selection criteria yield relatively isolated hosts and their
satellites, and it is worth noting that both the Milky Way and M31
would be rejected as host galaxies under our selection criteria.
We also note that, although we have adopted one particular
host--satellite selection algorithm, the results are not particularly
sensitive to the details of the selection algorithm (see, e.g., Brainerd 2005;
Agustsson \& Brainerd 2006, hereafter AB06).

We know from the MRS that the hosts will span a wide 
range of virial masses and, hence, a wide range of virial radii.  Therefore,
very different parts of the halos are probed by applying a fixed search
aperture of 500~kpc for the satellites.  The selection technique that we
have used is, however, fairly standard in the literature, has the advantage
that it is simple to implement, and does not depend on any specific {\it a priori}
assumption that the luminosity of a galaxy is correlated with its mass.
There are some indications from previous studies (e.g., Yang06) that the
satellite anisotropy may be a function of radius and we will explore this
in the following section.

In addition, the simple host--satellite selection criteria that we 
adopt allow, at least in principle, for ``multi--homed'' satellites.  That 
is, in principle a given satellite could be paired with more than one host.
In practice, we find that this occurs extremely rarely 
and the results we present below are completely unaffected by the presence
of mutli--homed satellites.  Also, it is true that the selection criteria
allow for the presence of galaxies with luminosities
$L_{\rm host}/6.25 \le L \le L_{\rm host}/2.5$ nearby to the host, and these
galaxies are not used in our analyzes.  In practice $\sim 48$\%
of the hosts have ``non--selected'' satellites nearby to them.  Of
the hosts that have non--selected satellites, the vast majority (77\%) 
have only one (54\%) or two (46\%) non--selected satellites.   Because
of this, we refer to our host galaxies as being merely ``relatively''
isolated.

\subsection{SDSS Host--Satellite Catalog}

In addition to selection criteria above, we require that the images of the
SDSS galaxies are not associated with obvious aberrations in the imaging 
(for which we performed a visual check). We also require that
the host galaxies are not
located close to a survey edge (i.e., the host must be
surrounded by spectroscopic targets from the SDSS, within the area
of interest).  We limit our study to the redshift range $0.01 \le
z \leq 0.15$, where the lower limit helps ensure that
the peculiar velocities do not dominate over the Hubble flow,
and the upper limit simply reflects the fact that very few hosts can be
found beyond this redshift.  After imposing all of our selection criteria,
our primary SDSS catalog consists of 
4,487 hosts and 7,399 satellites.  Note, however, that the size of the 
SDSS catalog is reduced when, below,
we further restrict our analyzes to SDSS
galaxies with measured stellar masses and specific star formation rates
(see Table~1).

\subsection{MRS Host--Satellite Catalog}

We select host and satellite galaxies from the mock redshift survey of
the MRS using the same 
redshift space proximity criteria that we
used for the SDSS.  Because of the shear size of the simulation, this 
results in a very large sample consisting of 70,882 hosts (of which
30\% are classified as elliptical) and 140,712 satellites.
In addition we note that the semi-analytic model of De Lucia \&
Blaizot (2007) distinguishes each MRS
galaxy according to three distinct types: type 0, type 1, and type 2.
Type 0 galaxies are the central galaxies of their 
friends-of-friends (FOF) halos.  These objects
are fed by radiative cooling from the surrounding halo.
Type 1 galaxies are the central galaxies of
``subhalos'', and they have their own self--bound dark matter subhalo.  Type 2
galaxies have been stripped of their dark matter and they lack distinct
substructure.  In our catalog of MRS
host--satellite pairs,
94\% of the hosts are the central galaxies of
their own FOF halo (i.e., they are type 0 objects).  This 
assures us that our prescription for finding host
galaxies is working well.  In contrast to the MRS hosts,
the MRS satellites are primarily type 1 objects (41\% of the sample)
or type 2 objects (39\% of the sample).  That is, the vast majority of the MRS 
satellites that are selected by proximity to the host in redshift space
are, indeed, contained within a larger halo.
However, 20\%  of the
MRS satellites are central galaxies of their
own FOF halo (i.e., they are type 0 objects).  These latter objects are
examples of ``interlopers'' --
objects which pass the redshift space proximity tests but which
are not necessarily
nearby to a host galaxy.  Without actual distance information for the
galaxies, a certain amount of interloper contamination of the satellite
population cannot be avoided.  However, since
the SDSS and MRS
host--satellite catalogs were selected in the same way, we expect
that the contamination of the SDSS satellite sample by interlopers will
be similar to that for the MRS sample.  We will investigate the effects
of interlopers on the observed locations of satellite galaxies in 
\S4.2 below.

A summary of the basic properties of the hosts and satellites in the SDSS
(left panels) and the MRS (right panels) is shown in Fig.~1.  From top
to bottom, the panels of Fig.~1 show probability distributions for the
number of satellites per host (panels a and b), 
the redshift distributions of the hosts (panels c and d), the
distribution of apparent magnitudes for the hosts
and satellites (panels e and f), the
distribution of absolute magnitudes for the hosts
and satellites (panels g and h), and the distribution of stellar
masses for the hosts and satellites (panels i and j).  Although
the distributions are not identical for the SDSS and MRS, they
are sufficiently similar that a direct comparison of the locations of the
satellites in the SDSS and the MRS should be meaningful.

One of the great luxuries of simulations (as opposed to observations
of the real Universe) is that all the information about the simulated 
galaxies is known.  In the remainder of this section we highlight some
of the information about the MRS hosts and satellites that, for
the most part, is not known for the SDSS hosts and satellites. 
Fig.~2 shows the relationship between the halo virial mass and the
stellar mass for the MRS hosts (left panel), the dependence of the halo
virial mass on absolute $r$--band magnitude for the MRS hosts
(middle panel), and the variation of stellar mass with 
$(g-r)$ for the MRS hosts (right panel).  
From Fig.~2, then, it is clear that the stellar mass of
the MRS hosts correlates well with the virial mass of the halo and, therefore,
the absolute magnitude.  In addition, it is clear that the reddest MRS host
galaxies are also the most massive hosts in the simulation. 

Fig.~3 highlights information that is known about the MRS satellites.
To construct this figure, we use only those objects
which we consider to be genuine satellites in the host--satellite catalog.
We make this restriction for Fig.~3 because here we
are interested in the properties of the genuine satellites, not the
properties of the interlopers.  Here we accept as genuine satellites those
objects that are located within a physical distance, $r_{3D} \le 500$~kpc,
of a host galaxy.  This is a rather non--restrictive definition of a 
genuine satellite and is based simply upon a match to the search radius
(i.e., $r_p \le 500$~kpc) that is used in our host--satellite selection
criteria (see \S3.1).  In addition we define the redshift at which the satellite
first enters its host's halo to be the redshift at which the satellite first
becomes a member of the FOF group of particles to which the host belongs.
The top panels of Fig.~3 show that the stellar masses
of the MRS satellites correlate well with the absolute magnitude (panel a),
$(g-r)$ color (panel b), and the redshift at which the satellites first entered
the halos of their hosts (panel c).  That is, the more massive is a satellite,
the more luminous is the satellite, the redder it is at the present day, and
the earlier it first entered the halo of its host.
This agrees well with the results of Kang07 from
their analysis of the redshifts at which satellite galaxies with various masses
and colors first entered the halos surrounding central galaxies in
group systems.   Fig.~3d) shows that there
is a strong correlation of the present--day color of a satellite and the
redshift at which it first entered its host's halo; the very reddest satellites
entered the halo more than 10~Gyr in the past, and the very bluest satellites
entered the halo within the past 1.5~Gyr.  Fig.~3f) shows that the projected
distance at which a satellite is found at the present day is also a strong function
of the redshift at which the satellite first entered the halo; on average,
satellites
at $r_p < 50$~kpc entered their hosts' halos $\sim 3.5$~Gyr earlier than satellites
at $r_p \sim 400$~kpc.  Fig.~3e) shows the ratio of the satellite to host stellar
mass as a function of the redshift at which the satellites first entered their
hosts' halos.  The majority of host--satellite pairs (84\%) have
mass ratios $\le 0.15$, and in the case of these pairs there is a monotonic
trend of mass ratio with $z_{\rm entry}$: the smaller is the mass ratio, the
more recently the satellite entered its hosts' halo.  In the case of the small
percentage of host--satellite pairs with mass ratios $> 0.15$, the trend
is reversed: the larger is the mass ratio, the more recently the satellite
entered its hosts' halo.

\section{Satellite Galaxy Locations: Analysis and Results}

The location of a satellite galaxy with respect to its host is
computed by measuring the angle, $\phi$, between the major axis of
the host and the direction vector on the sky that connects the
centroid of the satellite to the centroid of its host. 
Throughout we will refer to the angle $\phi$ as the ``location''
of the satellite.  Because we
are simply interested in investigating any preferential alignment
of the satellite locations with the semi--major axes of the hosts,
$\phi$ is restricted to the range [0$^\circ$, 90$^\circ$].
By definition, a value of $\phi = 0^{\circ}$ indicates alignment
with the host major axis, while a value of $\phi = 90^{\circ}$
indicates alignment with the host minor axis. 

Fig.~4 shows the probability distribution for the locations of the
satellite
galaxies in the SDSS (left panels) and the MRS (right panels) that were
selected using the redshift space proximity criteria from \S3.1.  In
this figure we have computed $\phi$ for all satellites and we have made
no subdivisions of the data based on host properties, satellite 
properties, or the projected distances at which the satellites are 
found.  The top
panels of Fig.~4 show the differential probability distributions, $P(\phi)$,
where the error bars have been computed from 1000 bootstrap resamplings
of the data.  Also shown in the top panels of Fig.~4 is the mean
satellite location, $\left< \phi \right> $, along with
the confidence levels at which
the $\chi^2$ test rejects uniform distributions for $P(\phi)$.
The bottom panels of Fig.~4
show the cumulative probability distributions for the satellite locations, 
$P(\phi \le \phi_{\rm max})$, along with the confidence levels at which the
Kolmogorov--Smirnov (KS) test rejects uniform distributions for 
$P(\phi \le \phi_{\rm max})$.  It is clear from Fig.~4 that the satellites
in both the SDSS and the MRS are located preferentially near the major
axes of their hosts, and the effect is detected with very high significance.
However, the tendency for satellites to be found near the major axes
of their hosts is stronger in the MRS than it is in the SDSS.
It is likely that
this discrepancy is due to the rather idealized way in which the
MRS host galaxies have been placed within their halos, and may point
to a modest misalignment between mass and light in the host galaxies
(e.g., AB06, Kang07, Bailin et al.\ 2005).

\subsection{Dependence of the Anisotropy on Host \& Satellite Properties}

In this subsection we explore ways in which the locations
of satellite galaxies may depend upon various physical properties of the
hosts and satellites. 
Fig.~5 shows results for the dependence of satellite location on 
various properties of the hosts.  Results for the SDSS satellites are
shown in the left panels of Fig.~5
and results for the MRS satellites are shown in the right panels.
The top panels of Fig.~5 show
the mean satellite location, $\left< \phi \right>$, as a function
of the host's $(g-r)$ color, computed at $z=0$.  In the case of the SDSS satellites,
$\left< \phi \right>$ is a strong function of host color, with the satellites
of the reddest MRS hosts exhibiting a large degree of anisotropy, while the satellites of
the bluest SDSS hosts are 
consistent with being distributed isotropically around their hosts.
In the case of the MRS satellites,
the satellites of red hosts are also
distributed much more anisotropically than are
the satellites of blue hosts. However, there is
also a clear anisotropy present in the locations of the satellites of the bluest
MRS hosts that is not seen for the satellites of the bluest SDSS hosts.

The middle panels of Fig.~5 show the dependence of $\left< \phi \right>$
on the specific star formation rate (SSFR) of the host.  
Here is it clear that in both the SDSS
and the MRS, the mean satellite location is a strong function of the
SSFR; the lower the SSFR, the more anisotropically distributed are the 
satellites. 
The bottom panels of Fig.~5 show the dependence of $\left< \phi \right>$
on the stellar mass of the host.  
From these panels, then, the mean locations of the satellites in 
both the SDSS and the MRS are functions of the stellar mass of the host;
the greater is the mass of the host, the more anisotropic are the locations
of the satellites. Overall,
the dependence of the mean satellite location, $\left< \phi
\right>$, on host color, SSFR, and stellar mass agrees fairly well between the
SDSS and MRS satellites.  While the precise values of $\left< \phi \right> $
are not identical in the two samples, a general trend is clear in both cases.
The satellites of hosts that are red, massive, and have low SSFR are distributed
much more anisotropically than are the satellites of hosts that are blue,
low mass, and have high SSFR.

Our results in Fig.~5e and Fig.~5f are somewhat at odds with the results of
Yang06 and Kang07 (i.e., we find that the
locations of the satellites are a function of the stellar mass of the host).  Yang06
found a weak tendency for the anisotropy in the locations of the
satellites of primary galaxies in SDSS group systems to
increase with the mass of the halos.  In particular, Yang06 found
that the mean location of the satellites of primaries with halo masses in the
range $1.4\times 10^{12}~ M_\odot \le M \le 1.4\times 10^{13}~ M_\odot$
was $\left< \phi \right> = 43.1^\circ\pm 0.4^\circ$ while the mean location
of the satellites of primaries with halo masses in the range
$1.4\times 10^{14}~ M_\odot \le M \le 1.4\times 10^{15}~ M_\odot$
was $\left< \phi \right> = 40.7^\circ\pm 0.5^\circ$.  That is, an increase in
the masses of the halos by a factor of $\sim 100$ resulted in a
decrease in $\left< \phi \right>$ of $2.4^\circ \pm 0.6^\circ$.  We 
find $\left< \phi \right> = 44.4^\circ \pm 0.6^\circ$ for the satellites of
relatively isolated SDSS hosts with $M_{\rm stellar} \sim 3\times 10^{10}~ M_\odot$
and $\left< \phi \right> = 41.3^\circ \pm 0.6^\circ$ for the satellites of
relatively isolated SDSS hosts with $M_{\rm stellar} \sim 3\times 10^{11}~ M_\odot$;
i.e., we see a decrease in $\left< \phi \right>$ of $3.1^\circ \pm 0.8^\circ$.
We do not know the masses of the halos of our SDSS hosts, but from Fig.~3a 
(i.e., the correlation of $M_{\rm stellar}$ with $M_{\rm virial}$ for 
the MRS hosts) we expect that this stellar mass range for our SDSS 
hosts corresponds to a 
factor of $\sim 30$ in halo mass.   Therefore, we see a similar decrease
in the value of $\left< \phi \right>$ in only $\sim 1$ order of magnitude in
mass for our sample as Yang06 saw in $\sim 2$ orders of magnitude in mass
for their sample.
Based on a simple extrapolation of our
results for the satellites of relatively isolated
SDSS hosts, we might therefore have expected the
satellites in the study of Yang06 to show a greater difference (by factor
of $\sim 3$ to 4) in the dependence of their locations on halo mass. 

In their simulation, Kang07 found no dependence of the satellite locations
on the masses of the halos that surrounded the primaries, and they explain that
this is due to the fact that the greater flattening of the higher mass halos
is counterbalanced by the satellites of lower mass halos having locations that
are somewhat flatter than the mass of the surrounding halo.
In an attempt to understand
the discrepancy between our results and those of
Kang07, we expand upon our result
for the dependence of the satellite locations on host mass in Fig.~6, where
we investigate the effects of the host image assignment prescription on
$\left< \phi \right>$ for the MRS galaxies.  The left panels of Fig.~6 show
results for MRS hosts that are classified as elliptical.  These are objects
for which the luminous galaxy is assumed to share the shape of the surrounding
dark matter halo.  The right panels of Fig.~6 show results for MRS hosts
that are classified as non--elliptical.  These are objects for which the
luminous galaxy is assumed to be a thin disk, oriented such that the angular
momentum of the disk aligns with the net angular momentum of the surrounding
halo.  The top panels of Fig.~6 show $\left< \phi \right>$ as a function of
host color.  From these panels, it is clear that the satellite anisotropy is
stronger for the very reddest elliptical MRS hosts than it is for the bluest
elliptical MRS hosts, however there is essentially no dependence on host color
for the locations of the satellites of non--elliptical MRS hosts.  It is also
clear that the satellites of the elliptical MRS hosts show a much greater
degree of anisotropy in their locations compared to the satellites of
non--elliptical MRS hosts.  This is due to the fact that strict alignment of
mass and light in the numerical galaxies, as was done for the elliptical 
MRS hosts, maximizes the anisotropy of the
satellite locations (see AB06 and Kang07).

The bottom panels of Fig.~6 show the dependence of the satellite locations
on the stellar masses of the MRS hosts.  From these panels
it is clear that, at fixed host mass, the satellites of the elliptical
MRS hosts show a greater degree of anisotropy in their locations than do the
satellites of non--elliptical MRS hosts.  In addition,
{\it within a given
class of MRS host galaxy} there is no clear trend of $\left< \phi \right>$ with
the stellar mass of the host.  That is, the trend with host stellar mass that
we see in panel f) of Fig.~5 is due to the fact that the lowest mass MRS hosts
are non--ellipticals (whose satellites show a relatively small degree of
anisotropy in their locations) while the highest mass MRS hosts are ellipticals
(whose satellites show a much greater anisotropy in their locations).  The fact
that, within a particular image assignment prescription for the MRS hosts, 
we see no dependence of
$\left< \phi \right>$ on host mass probably explains why Kang07 did not see
a strong dependence of the satellite anisotropy on the masses of the central
galaxies in their simulation.  Kang07 did not assign galaxy types to
their central galaxies, and they used the same prescription to assign image shapes
to all the luminous galaxies in their simulation.  

In Fig.~7 we demonstrate the effect on $\left< \phi \right>$
if we use the same image assignment 
scheme for all of the MRS hosts.  That is, Fig.~7 shows how $\left< \phi \right>$ is
affected if we do not adjust our image assignment scheme according to whether or
not the MRS host galaxy is an ``elliptical'' or a ``non--elliptical''.  Open triangles 
in Fig.~7 show the dependence of $\left< \phi \right>$ on host color (left panel),
SSFR (middle panel), and stellar mass (right panel) under the assumption that all
MRS host galaxies share the shapes of their dark matter halos.  That is, the open triangles
in this figure show the resulting values of $\left< \phi \right>$ if we simply apply
the ``elliptical'' image assignment scheme to all MRS hosts.  Open circles in
Fig.~7 show the result of simply applying the ``non--elliptical'' image assignment
scheme to all MRS hosts.  That is, the open circles show the result that occurs if
all MRS hosts are assumed to be thin disks, oriented such that the angular momentum
of the disk is perfectly aligned with the net angular momentum of the halo.  For
comparison, solid squares show the results from Fig.~5 for the SDSS hosts and
satellites.  From Fig.~7, then,
if we adopt the same image assignment scheme for all MRS hosts, independent
of their bulge--to--disk ratios, we cannot 
reproduce the observed dependence of $\left< \phi \right>$ on host color, SSFR,
and stellar mass that we find for the SDSS galaxies.
If we use a single image assignment scheme for all MRS hosts, 
$\left< \phi \right>$ for the MRS satellites
generally has a much weaker dependence on host color, SSFR, and stellar mass than we see
in the SDSS, and sometimes the dependence of $\left< \phi \right>$ on host property
is actually opposite to what we see in the SDSS.  Fig.~7 then argues rather strongly
for the need for two distinct image assignment schemes as we have adopted for the
elliptical and non--elliptical MRS hosts.  It also suggests that luminous elliptical 
galaxies and luminous spiral galaxies in the observed Universe are oriented within
their dark matter halos in rather different ways.

In Figs.~8 through 10 we expand upon our results in Fig.~5 for the
dependence of the satellite locations on host color, 
and we do this by splitting
our sample into ``red'' hosts and ``blue'' hosts.
To define ``red'' and ``blue'', we fit
the distributions of $(g-r)$ host colors in the top panels of Fig.~5
by the sum of two Gaussians (e.g., Strateva et al.\ 2001; Weinmann et al.\
2006).   We find that the division between
the two Gaussians lies at $(g-r) = 0.7$ for the SDSS galaxies and
at $(g-r) = 0.75$ for the
MRS galaxies.  We therefore define SDSS
hosts with $(g-r) < 0.7$ to be ``blue'' and SDSS hosts with $(g-r) \ge
0.7$ to be ``red''.  Similarly, we define MRS hosts with $(g-r) < 0.75$ to
be ``blue'' and MRS hosts with $(g-r) \ge 0.75$ to be ``red''.
Figs.~8 and 9 then show $P(\phi)$ and $P(\phi \le
\phi_{\rm max})$ for satellites of the red and blue hosts, respectively.

It is clear from Figs.~8 and 9 that the satellites of red hosts
have a much stronger preference for being located near the major axes
of their hosts than do the satellites of blue hosts.  This is true for 
both the SDSS and MRS satellites.  In addition,
the MRS satellites show a stronger
preference for being located near the major axes of their hosts than do the
SDSS satellites.  The satellites of blue SDSS hosts
are consistent with having an isotropic distribution around their hosts, while
the satellites of red SDSS hosts have a strong preference for being located
near the major axes of their hosts.
Such a disparity in the locations of the satellites of red and blue host
galaxies was also found by APPZ, Kang07, Yang06, Bailin et al.\ (2008), 
and Siverd et al.\ (2009),
with the satellites of blue hosts showing little to no preference
for a particular location relative to their hosts.

In the case of APPZ, small number statistics (i.e., a relatively 
small number of host--satellite pairs in these studies) 
prevented them from placing a strong
constraint on whether or not the locations of the satellites of blue hosts
were, in fact,  truly different from the locations of
the satellites of the red hosts.   The cause of this is
two--fold.  First, 
the majority of SDSS hosts are red (see Table~1).  
Second, the blue hosts
tend to have fewer satellites
than do their red counterparts.  This results in a paucity of host-satellite pairs in 
which the host is blue.  Here, however,
our sample of SDSS hosts
and satellites is sufficiently large that we can make a definitive 
statement about the locations of the satellites of blue hosts versus
the locations of the satellites of red hosts.  
To do this, we computed a two--sample KS test using the 
cumulative probability distributions from the bottom left panels of Figs.~8
and 9.  The result is that, at the 99.9\% confidence level,
the KS test rejects the null hypothesis that
the locations of the satellites of red SDSS hosts are drawn from the same
distribution as the locations of the satellites of blue SDSS hosts.  That
is, with high significance, the locations of the satellites of red and blue
SDSS hosts are truly different.

Fig.~10 illustrates
the underlying cause of the ``lack'' of anisotropy in the locations of
the satellites of the blue SDSS hosts.  Here we plot the
mean satellite location, $\left< \phi \right>$, as a function of 
projected distance.  The left panels of Fig.~10 show the results for the
satellites of red hosts, while the right panels show the results for the
satellites of blue hosts.  In the case of the satellites of red hosts, $\left<
\phi \right>$ is largely independent of $r_p$.  Hence, when we average the
satellite locations over all projected distances, $r_p \le 500$~kpc (i.e.,
as in Figs.~5, 8 and 9), the result is that the satellites of red hosts exhibit
a strong degree of anisotropy.  In the case of the satellites of blue hosts,
however, $\left< \phi \right>$ is a function of $r_p$.  Satellites of
blue hosts that are located at small projected distances have a tendency
to be found close to the
major axes of their hosts, while satellites of blue hosts with larger projected 
distances exhibit a different degree of anisotropy.  In particular,
satellites of blue SDSS hosts that have large values of $r_p$  have a tendency
to be found close to the {\it minor} axes of their hosts, and when
the locations of all satellites of the blue SDSS hosts are averaged
over all projected distances, $r_p \le 500$~kpc, the result is consistent with
an isotropic distribution (i.e., top left panel of Fig.~9).  The satellites
of blue MRS hosts show a preference for being located close to the major axes
of their hosts for projected distances $r_p < 300$~kpc, but at larger 
projected distances the satellite locations become consistent with a random
distribution.  Therefore,
the net anisotropy of the MRS satellites of blue hosts
is substantially reduced when averaged over all
values of $r_p \le 500$~kpc (i.e., top right panel of Fig.~9).

Fig.~11 shows the dependence of the mean satellite location as
a function of various properties of the satellites.
Panels a) and b) show the dependence of $\left< \phi \right>$ on
$(g-r)$, panels c) and d) show the dependence of $\left< \phi 
\right>$ on specific star formation rate, panels e) and f) show the
dependence of $\left< \phi \right>$ on the stellar mass, and
panels g) and h) show the dependence of $\left< \phi \right>$ on the 
projected distances at which the satellites are found.  
As in Fig.~5, there is generally
good agreement between the results for SDSS satellites (left panels) and
MRS satellites (right panels), 
with the greatest degree of anisotropy being shown
by the reddest, most massive, and lowest--SSFR satellites.  The 
locations of the bluest, least massive, and highest--SSFR satellites show
little to no anisotropy.  This is in part attributable to the fact that
these objects are likely to have been accreted in the very
recent past (see, e.g.,
Fig.~3); however, as we will see in the next section this is also partially
attributable to the fact that our blue satellite population is heavily
contaminated with interlopers whose effect is to strongly suppress the anisotropy.

Finally, we note
that the locations of the satellites are weakly--dependent upon the
projected distances at which they are found (panels g and h of Fig.~11), with
the satellites found at $r_p \sim 450$~kpc showing less anisotropy than
satellites found at smaller projected distances.  This is, of course,
unsurprising since the objects that are found
at large $r_p$ are most likely to be either genuine satellites that have been
accreted very recently (see, e.g., panel f of Fig.~3) or 
interlopers.  In addition, we note that, contrary to the claims
of Bailin et al.\ (2008) very few of our SDSS satellites are found at
projected distances $r_p < 50$~kpc (see the histogram in 
Fig.~5g).  The lack of SDSS satellites
at small projected distances is caused primarily by the fact
that fiber collisions prevent the simultaneous measurement of the redshifts of
two galaxies that are very close to each other on the sky.  So, it is only in regions
of the sky that were observed multiple times that satellites with small values
of $r_p$ may be found.  Also, because we have performed a visual check
of each and every host galaxy, we know for certain that the satellites that
we do identify at $r_p < 50$~kpc are, indeed, separate from their host.  That is,
the satellites at these projected distances are not, say, H-II regions or
bright blue knots within the host galaxy
that have been misidentified as objects that are distinct from the host galaxy.

\subsection{Effects
of Interlopers and $z_{\rm entry}$}

When discussing the satellites, it is important to remember that 
at least some fraction
of the satellites that are found using the selection criteria in \S3.1 are not
genuine satellites at all.  Rather, they are interlopers that are not
necessarily nearby to a host galaxy, but they happen to pass all of
the proximity and magnitude criteria in order 
to be included as satellites in the
catalog.  In the case of the SDSS satellites, we have no way of knowing which
of the satellites in our catalog are real and which are interlopers.  In the
case of the MRS satellites, however, we have full phase--space information
and we know the physical distances of each of the satellites in the 
catalog from their respective hosts.
Until now, all of our
calculations of the locations of satellite galaxies in the MRS have included
both the satellites that are physically close to host galaxies, as well
as the interlopers.  This was done in order to
better compare the MRS to the SDSS via identical procedures for the 
identification of hosts and satellites. In this section 
we will examine the effects of the interlopers on the observed
anisotropic distribution of the satellites, as well as the effect of
the redshift at which the satellites first entered their hosts' halos.

Here we adopt the same
rather non--restrictive definition of a genuine satellite as in \S3.3 and we
accept as genuine
satellites those objects that are located within a physical distance
$r_{3D} \le 500$~kpc of a host galaxy.  The mean location of all MRS satellites
that are found within $r_{3D} \le 500$~kpc of a host galaxy is
$\left< \phi \right> = 39.12^\circ \pm 0.08^\circ$, while the mean location
of the interlopers is $\left< \phi \right> = 43.6^\circ \pm 0.1^\circ$.  Clearly,
then, the presence of the interlopers in the full data set reduces the
measured anisotropy in the satellite locations compared to what one would
measure in the absence of the interlopers.  Interestingly, the interlopers are
not randomly--distributed around the hosts.  Instead, on average the interlopers
show a weak preference for being located near the major axes of the hosts.
This is due to the fact that relatively
few interlopers are located at extremely large distances from the host galaxies.
The median distance of the interlopers from the hosts is only 630~kpc, indicating
that by and large they are within the local vicinity of the hosts.  

Shown in Fig.~12 are the results for the differential probability 
distribution, $P(\phi)$, for MRS satellites, with and without the 
contribution of interlopers.  
The open points in Fig.~12 show
$P(\phi)$, computed using all satellites in the MRS catalog, including
the interlopers.  
The filled points show $P(\phi)$, computed using only
the satellites in the MRS catalog that are located within
a physical distance $r_{3D} \le 500$~kpc of their host.  
Included in 
each of the panels of Fig.~12 is the value of the mean satellite location,
with and without the contribution of interlopers, along with the
fraction of satellites in the MRS catalog that are interlopers 
(i.e., objects which have $r_{3D} > 500$~kpc).
As above, the net effect of interlopers is to reduce
the value of $\left< \phi \right>$.
The top panels of Fig.~12 show $\left< \phi \right>$ for the satellites
of red MRS hosts (left panel) and the satellites of blue MRS hosts
(right panel).  The fraction of interlopers 
is nearly identical;
interlopers account for 32\% of the satellites of red MRS hosts
and 35\% of the satellites of blue MRS hosts.
The presence of the interlopers reduces $\left< \phi \right>$ by similar
amounts for the satellites of both the red and blue MRS hosts.  

We note
that the presence of interlopers is not the cause of the reduced anisotropy 
for the satellites of the blue hosts compared to the satellites of the
red hosts.  That is, the removal of the interlopers from the MRS sample
does not result in the locations of the satellites of blue MRS hosts being
the same as those of red MRS hosts.  Formally, when the interlopers
are removed, the mean location of the
MRS satellites surrounding blue hosts differs from the mean location of the
MRS satellites surrounding red hosts by more than $20\sigma$.  This 
differs from the conclusions
of Kang07 who found that removing the interlopers from their sample resulted
in the locations of the satellites of blue central galaxies being the
same as the locations of the satellites of red central galaxies.
However, as with the dependence of satellite anisotropy on
host mass, this difference may be simply attributable to the two
different prescriptions that we
have used to assign images to the luminous MRS host galaxies.  That is,
on average, the red MRS hosts are ellipticals and the blue MRS hosts
are non--ellipticals.  From Fig.~6, then, we would automatically expect 
the satellites of red MRS hosts to show a greater degree of anisotropy
in their locations than the satellites of blue MRS hosts because of the
strong correlation of the satellite anisotropy with the 
host image assignment scheme (i.e., our ``elliptical'' image assignment
scheme maximizes the satellite anisotropy).

The bottom panels of Fig.~12 show $\left< \phi \right>$ for red
MRS satellites (left panel) and blue MRS satellites (right panel).
Here the interloper fraction is strikingly different; only 19\% 
of the red MRS satellites are interlopers,
while 57\% of the blue MRS satellites are interlopers.
Therefore, the presence of a large number of interlopers in the sample of
blue satellites is a major factor in the reduced anisotropy of blue 
satellites compared to red satellites (e.g., panels a) and b) of Fig.~11).

As noted by Kang07, the redshift at which a genuine satellite
first enters the halo of its host is a strong function of the mass of
the satellite and the present--day color of the satellite.  From panels c)
and d) of Fig.~3, the more massive
the satellite and the redder
is its present--day $(g-r)$ color, the earlier the satellite
made its first entry into
the halo of its host (see also Kang07).  
One would naturally expect that it would take a few crossing
times for satellites to have their trajectories affected to the point
where the locations of the satellites would provide a good
proxy for the distribution of the mass with the host's halo.  For a CDM
halo with a mass of $\sim 10^{12} M_\odot$ and virial 
radius $\sim 180h^{-1}$~kpc,
the crossing time will be of order $\tau_{\rm cross} \simeq
R/v \simeq 1.7$~Gyr for $v \sim 150$~km~sec$^{-1}$.  Therefore, unless 
the infall of satellites is highly non--spherical, we would expect
satellites that arrived within their host's halo within the
past billion years should show markedly less anisotropy than satellites
that arrived within their host's halo in the much more distant past.

Solid squares in the top panel of Fig.~13 show the mean satellite location, $\left<
\phi \right>$,  as a function of the redshift at which the genuine MRS 
satellites first entered their hosts' halos.  From this figure, 
satellites that first entered their host's halo within the past $\sim 1.25$~Gyr
(i.e., $z_{\rm entry} \sim 0.1$) show
considerably less anisotropy than do those which first entered their host's
halo at earlier times. 
Referring to the bottom left panel of Fig.~3, the bluest
MRS satellites are those which
first entered their host's halo at redshifts $z_{\rm entry} \sim 0.1$, while the
reddest MRS satellites are those which first entered their
host's halo at redshifts $z_{\rm entry} > 2$.  Therefore,
it is unsurprising that, after the removal of interlopers with 
$r_{3D} > 500$~kpc,
the degree of anisotropy exhibited by the 
blue MRS satellites (bottom right panel of Fig.~12, $\left< \phi \right> =
41.9^\circ \pm 0.2^\circ$) is considerably
less than the degree of anisotropy exhibited by the red genuine MRS
satellites (bottom left panel of Fig.~12, $\left< \phi \right> = 
38.2^\circ \pm 0.1^\circ$).  
Also shown in the top panel
of Fig.~13 is the mean satellite location, $\left< \phi \right>$,
as a function of $z_{\rm entry}$ 
for the genuine satellites of red MRS hosts (open triangles) and 
the the genuine satellites of blue MRS hosts (open circles).  From this
figure, then, it is clear that satellites began arriving within the halos
of the red MRS hosts much earlier than did the satellites of blue MRS hosts.
The bottom panel of Fig.~13 shows the probability of the entry redshift,
$P(z_{\rm entry})$, for the type 1 and type 2 MRS satellites.
The type 2 satellites are the objects
that have been stripped of their dark matter and, as expected, Fig.~13
shows that $z_{\rm entry}$ is, on average, considerably earlier for the type~2
satellites than it is for the type 1 satellites (which still retain
their dark matter).

\section{Summary and Comparison to Previous Results}

Here we summarize the major results of our study and compare them to 
results of previous, similar investigations.  The major results
that we have obtained by computing the mean satellite location,
$\left< \phi \right>$, using all satellites (including interlopers) are: 

\begin{itemize}
\item[1.] $\left< \phi \right>$ is a function of the host color,
specific star formation rate, and stellar mass.  Satellites of red, massive
hosts with low SSFR show considerably more anisotropy than do satellites of
blue, low mass hosts with high SSFR
(Fig.~5).

\item[2.] In order to reproduce the observed trends for the dependence of
$\left< \phi \right>$ on host color, SSFR, and stellar mass,  we require two
distinct image assignment prescriptions for the simulated galaxies: ellipticals
share the shapes of their dark matter halos and non--ellipticals have their
angular momentum vectors aligned with the net angular momentum of the halo.
(Fig.~7)

\item[3.] $\left< \phi \right>$ is a function of the satellite color,
specific star formation rate, and stellar mass.  Red, massive satellites with
low SSFR show considerably more anisotropy than do blue, low mass satellites
with high SSFR (Fig.~11).

\item[4.] Averaged over all satellites at all
projected distances, the locations of the satellites of blue SDSS host galaxies
are consistent with
an isotropic distribution, while the satellites of red
SDSS host galaxies have a strong preference for being found near the major
axes of their hosts.  At the 99.9\% confidence level, the two 
distributions are inconsistent with having
been drawn from the same parent distribution (Figs.~8 and 9).

\item[5.] Satellites of blue MRS host galaxies are found preferentially
close to the major axes of their hosts, however the degree of anisotropy
is considerably less than that shown by the satellites of red MRS
host galaxies (Figs.~8 and 9).

\item[6.] $\left< \phi \right>$ for the satellites
of red host galaxies is approximately independent of $r_p$, while
$\left< \phi \right>$ for the satellites of blue host galaxies is an
increasing function of $r_p$  (Fig.~10).

\end{itemize}

\noindent
The major results that we have obtained with regards to interlopers
are:

\begin{itemize}
\item[7.]
The interloper contamination is similar (32\% and 35\%, respectively) for the satellites
of red MRS hosts and blue MRS hosts (Fig.~12, top panels).

\item[8.] Interlopers are not the cause of the different amount of anisotropy
shown by the locations of the satellites of blue MRS hosts versus 
the satellites of red MRS hosts.  
The genuine satellites
of red MRS hosts show considerably more anisotropy than do the genuine
satellites of blue MRS hosts, and the significance is 
greater than $20\sigma$ (Fig.~12, top panels). 

\item[9.] Our host--satellite selection criteria result in
57\% of the blue satellites in the MRS catalog being interlopers and
19\% of the red satellites being interlopers
(Fig.~12, bottom panels).

\item[10.] At the 16$\sigma$ level, the red genuine MRS satellites show
considerably more anisotropy in their locations than do the blue genuine
MRS satellites  (Fig.~12, bottom panels).  This is due to the fact that
the blue satellites have only recently arrived within their hosts' halos, 
while the red satellites arrived in the far distant past.

\end{itemize}

As mentioned above, the general trend for the satellites of red hosts
to show considerably more anisotropy than those of blue hosts has
been observed by others (e.g., APPZ; Yang06; Kang07; Bailin et al.\ 2008;
Siverd et al.\ 2009),
and our results agree well with these previous results.  Further, 
we have demonstrated conclusively that in the case of relatively isolated 
host--satellite systems, the satellites of blue host galaxies are distributed
differently around their hosts than are the satellites of red host galaxies.

Also as mentioned above, although our results for the satellites of  SDSS 
host galaxies show trends that are very similar to our results for the satellites
of MRS host galaxies,  the satellites of MRS host galaxies exhibit a greater
degree of anisotropy in their locations.  This is probably attributable
to the simple prescriptions that we have used to define the images 
of the MRS host galaxies, and may indicate that a certain degree of
misalignment of the galaxy images from our idealized prescriptions
is necessary
(see also AB06; Kang07; Okumura et al.\ 2009;
Faltenbacher et al.\ 2009; Okumura \& Jing 2009). 
To estimate the degree of misalignment that is necessary
for the anisotropy of the locations of the
satellites of the MRS galaxies to match those of the SDSS galaxies, we add
Gaussian--random errors to the orientations of the MRS host galaxy images (as
viewed in projection on the sky).  When we do this, we find that a mean
misalignment of $|\delta \theta| \sim 20^\circ$ (measured relative to the
``idealized'' MRS host image)  reduces the anisotropy in
the locations of the satellites of the MRS hosts to the point that,
when averaged over $r_p \le 500$~kpc, the result agrees with the result for
the satellites of SDSS hosts.   We note that, although we have phrased
this in terms of a misalignment of the host galaxy image from the idealized
prescription, this should not be strictly interpreted as the mass and light of the
SDSS galaxies being misaligned by an average of
$\sim 20^\circ$.  While there may be some
degree of true misalignment, it is always important to keep in mind that 
there are observational errors associated with the measurement of the 
position angles of observed galaxies, and these can be particularly large
in the case of very round galaxies, or galaxies with well--resolved spiral
arms.  Such errors in the determination of the position
angles of the SDSS galaxies will, therefore, contribute some amount
to a need for 
misalignment of the host images in the MRS in order to match
the observations.  Unfortunately, errors for the position angles of the
SDSS galaxies are not yet available in the data base, so we are unable to
estimate the contribution of position angle errors to the value of
$|\delta \theta |$ above.

Although our work is very similar in spirit to that of Kang07, we arrive
at some different conclusions.  First, we find that the degree
of anisotropy in the satellite locations depends upon the
stellar mass of the host galaxy, while Kang07 found no dependence of
the satellite locations on the mass of the surrounding halo.
The discrepancy between our theoretical results and the
theoretical results of Kang07 is probably due to the fact that
we have chosen to use two different image assignment schemes for the MRS
hosts (ellipticals vs.\ non--ellipticals), while Kang07 use the same
image assignment scheme for all of their central galaxies.  We find that
{\it within a given image assignment scheme} there is no dependence of 
$\left< \phi \right>$ on host mass; however, there is considerably more
anisotropy shown by the satellites of elliptical MRS hosts than non--elliptical
MRS hosts.  This, combined with the fact that the least massive MRS hosts
are non--ellipticals and the most massive MRS hosts are ellipticals leads
to the trend of satellite anisotropy with host mass that we see in the simulation.

In their study of the locations of satellites in SDSS group systems, Yang06 found
a rather weak dependence of satellite location on the mass of the
surrounding halo; over two orders of magnitude in halo mass, the value of
$\left< \phi \right>$ decreased by only $2.4^\circ \pm 0.6^\circ$.  
By contrast, we appear to find a somewhat stronger trend of 
satellite location with host mass.  Over $\sim 1$~order of
magnitude in host mass we find a decrease in the value of $\left< \phi \right>$
that is similar to the value found by Yang06: $3.1^\circ \pm 0.8^\circ$.
A simple extrapolation of our results to much higher masses would suggest that
over the mass range of their sample, Yang06 should have found a greater
change in $\left< \phi \right>$.
The resolution of this discrepancy is unclear, but it could 
have to do with the fact
that we are investigating somewhat different systems (i.e., relatively isolated
hosts vs.\ group environments, where perhaps the central galaxy is not located
precisely at the dynamical center).   In addition, we use stellar masses to define the
masses of our host galaxies while Yang06 derive masses for the halos of their
groups using a conditional luminosity function.  This discrepancy
certainly warrants
further investigation in the future, particularly since $\Lambda$CDM 
predicts that the flattening of the dark matter halos of galaxies should
increase with halo virial mass (e.g.,  Warren et al.\ 1992; Jing \& Suto 2002;
Bailin \& Steinmetz 2005; Kasun \& Evrard 2005; Allgood et al.\ 2006). 

Additionally, in their simulation
Kang07 find that the reason the satellites of blue
central galaxies show less anisotropy than the satellites of red central
galaxies is that the presence of a large number of interlopers around the
blue central galaxies 
suppresses the anisotropy.  This is because Kang07 find
that there is a considerably larger number of interlopers in the sample 
of satellites around blue central galaxies ($\sim 35$\%) than there are
in the sample of satellites around red central galaxies ($\sim 15$\%).  
When Kang07 remove the interlopers, they find that the degree of
anisotropy shown by the genuine satellites of red and blue centrals is 
identical.  In our work we find a nearly identical interloper fraction 
for the satellites of red and blue host galaxies (32\% for red hosts 
and 35\% for blue hosts).   However, it is important to note that
we have used a simple non--iterative
technique to identify host and satellite galaxies,
while Kang07 use a sophisticated,  iterative technique which is supposed
to reduce the number of interlopers on average.  So, it is unsurprising
that our relative number of interlopers would differ.

When we remove the interlopers from the MRS host and satellite catalog, we find
that the satellites of blue hosts still show much less anisotropy
than do the satellites of red hosts.  In our analysis, there appear
to be two causes of the differences between the locations of the 
satellites of red and blue hosts.  First, $\left< \phi \right>$ is
largely independent of $r_p$ for the satellites of red hosts.  Therefore,
when $\left< \phi \right>$ is averaged over all projected distances,
$r_p \le 500$~kpc, the satellites of the red hosts show a great deal
of anisotropy.  In contrast, $\left< \phi \right>$ for the satellites
of blue hosts is a function of $r_p$, with satellites located at small
$r_p$ being found near the major axes of their hosts and satellites
located at larger distances having different locations (nearly isotropic
in the case of the MRS satellites, and near the minor axes of the hosts
in the case of the SDSS satellites).  Therefore, when $\left< \phi \right>$
is averaged over all projected distances, $r_p \le 500$~kpc, the satellites
of blue hosts show a markedly reduced anisotropy.  In addition, we know that
the blue MRS hosts are by and
large disk systems (``non--ellipticals'') and the satellites of the
non--elliptical MRS hosts are distributed much less anisotropically than are the
satellites of the elliptical MRS hosts due to our
image assignment schemes.  Thus, as with the discrepancy
regarding the trend of satellite anisotropy with host mass, the discrepancy
between our results and those of Kang07 for the origin of the different
amount of anisotropy shown by satellites of red and blue hosts may be due
in large part to the two different assignment schemes that we have used to
define the images of the MRS host galaxies.

Now, it is, of course, extremely important not to put too much significance on one
data point, especially in the case of a figure in which the data points
are inherently correlated.  Nevertheless, the value of $\left< \phi \right>$
for the satellites of blue SDSS hosts that are located at $r_p \sim 400$~kpc is
intriguing because it suggests a ``reversal'' of the anisotropy signal at
large distances (right panel of Fig.~10).  
In their sample of extremely isolated SDSS host galaxies
(much more isolated than our sample), Bailin
et al.\ (2008) found no statistically--significant
dependence of $\left< \phi \right>$ on $r_p$; however, their sample size is
much smaller than we have used here (337 hosts and 388 satellites).
A weak tendency for the satellites of isolated disk galaxies
to be aligned with the minor axes of the hosts was seen 
by Zaritsky et al.\ (1997) when the satellite locations were averaged out
to large projected distances ($r_p \sim 500$~kpc).  More recently,
Siverd et al.\ (2009) found a weak tendency for extremely faint satellites
of highly--inclined blue SDSS galaxies to have a minor axis preference when
the locations of the satellites were averaged out to similarly large 
projected distances.  This is tantalizing in light of the results
of Zhang et al.\ (2009) who found that the spin axes of dark matter halos
with mass $\lesssim 10^{13} M_\odot$ tend to be aligned along the filament
in which the halo resides.   In addition, Bailin et al.\ (2008) found
that satellites that are most likely to have been accreted recently
have a tendency to be found along the same axis as the large--scale
structure that surrounds the host galaxy.
Thus, a ``reversal'' of the anisotropy for
the locations of the satellites of disk host galaxies at large projected 
distances could indicate preferential infall of satellites along filaments.
Establishing the existence of
such a reversal of the anisotropy at large projected
distance will, of course, take a great deal more effort (see, e.g., Siverd
et al. 2009 who conclude that the discrepancies between previous investigations
are largely attributable to sample selection).

We have shown that satellites that are very blue, have low masses
and high SSFR tend to show little to no anisotropy in their locations while
satellites that are very red, have high masses and low SSFR show a great
deal of anisotropy in their locations.  Similar results have been seen Yang06,
Kang07, and Siverd et al.\ (2009).  Using their simulation, Kang07 interpret
this effect to be due to the fact that the reddest, most massive
satellites are those which entered their hosts' halos in the far distant past,
while the bluest, least massive satellites have only recently arrived within
the halo.  Our work with the MRS hosts and satellites
directly supports this conclusion, however there is an additional component to
the effect in our case.  The redshift space selection criteria that we have
adopted result in the majority of blue satellites (57\%) being interlopers,
the presence of which
reduces the anisotropy exhibited by the genuine blue satellites by a substantial
amount (a $7\sigma$ effect; see the bottom right panel of Fig.~12). 

It is, of course, a tremendous simplification to use the global
dark matter halo properties to obtain properties of the luminous central galaxy
as we have done here.  This is due to the fact that the scale size of
the luminous galaxy is far smaller than that of the halo in which it resides.
Therefore, it is not necessarily the case that the net halo shape or net halo
momentum will be reflected in the shape or angular momentum of the central
galaxy.  Given these caveats, it is really quite
remarkable that such naive prescriptions as we have adopted here give 
rise to a fair agreement between theory and observation.  If nothing
else, our results lend credence to the idea that large luminous galaxies
have some knowledge of the halo in which they reside, despite the fact that
the luminous galaxy may be an order of magnitude smaller in extent than its
dark matter halo.  While mass may not directly trace light within galaxies,
it would not be possible to have such similar results for the locations of
satellite galaxies in the observed Universe and $\Lambda$CDM if mass and 
light were not strongly coupled within the host galaxies.

\section{Conclusions}

Here we have shown that the locations of the
satellites of relatively isolated host galaxies in the
SDSS and the Millennium Run simulation (MRS) show very similar trends,
provided that we adopt two distinct image assignment prescriptions for
the MRS hosts: elliptical hosts share the shapes of their dark matter
halos while non--elliptical hosts have their angular momentum vectors
aligned with the net angular momentum of their halos.  If we use only
a single image assignment prescription for all MRS hosts, it is not
possible to reproduce the dependencies of the mean satellite location
on host properties that we see in the SDSS.
Averaged over all projected distances, $r_p$, the degree to
which satellites are found preferentially close to the major axes of
their hosts is a function of the host's stellar
mass, SSFR, and $(g-r)$ color.  The satellites of red, massive hosts
with low SSFR show a strong tendency for being located near the
major axes of their hosts, while the satellites of blue, low--mass hosts
with high SSFR show little to no anisotropy in their locations.  
Red, massive satellites with
low SSFR show a strong tendency for being located near the 
major axes of their hosts, while blue, low--mass satellites with high
SSFR show little to no anisotropy in their locations.  
This last trend can be understood in part by the
different times at which satellites entered their hosts' halos.  That is,
redder, more massive satellites entered their hosts' halos in the far distant
past while bluer, less massive satellites have only recently entered
their hosts' halos. Therefore, the blue satellites have had their kinematics affected 
less by their hosts than have the red satellites.  In the case of the 
blue satellites, however, there is an additional factor that reduces
the observed anisotropy.  From our analysis
of the MRS, we expect that the majority of the blue satellites are interlopers,
not genuine satellites, and the presence of these
objects greatly suppress the value of the measured anisotropy in
comparison to the intrinsic anisotropy.

Overall, the presence of interlopers in the satellite catalogs suppresses
the degree to which the satellites exhibit an anisotropy in their
locations.  However, even after the removal of the interlopers from
the catalog of MRS satellites, the satellites of blue MRS host
galaxies show substantially less anisotropy in their locations than
do the satellites of red MRS host galaxies.    There are two causes
for the reduction of the anisotropy for the satellites of blue hosts
versus the satellites of red hosts.  First, there is a marked difference
of the dependence of the mean satellite location on projected distance for 
the satellites of red hosts compared to the satellites of blue hosts.
In the case of the red SDSS and MRS hosts, the locations of the satellites are largely
independent of the projected distances at which they are found.
In the case of the satellites of blue SDSS hosts, we find that at large
projected distances ($r_p \sim 400$~kpc), there is a tendency for the satellites
to be found close to the {\it minor}  axes of their hosts, while at
smaller projected distances ($r_p \sim 100$~kpc) the satellites have a tendency
to be found close to the major axes of their hosts.  The satellites
of the blue MRS hosts that are found at small projected distances are located 
preferentially close to the major axes of the hosts, while at large projected
distances the locations of the satellites are essentially isotropic.  Therefore,
when the locations of the satellites of blue host galaxies are averaged over
all projected distances ($r_p \le 500$~kpc) there is a substantial reduction 
in the signal compared to when the locations of the satellites of red host
galaxies are averaged over all projected distances.

In addition, we find that the prescriptions we use to assign images
to the MRS host galaxies give rise to different degrees of anisotropy in the
satellite locations. 
The satellites of elliptical MRS hosts are distributed
much more anisotropically than are the satellites of non--elliptical MRS 
hosts.  Further, the red MRS hosts are by and large ellipticals, while the
blue MRS hosts are by and large non--ellipticals.  Therefore, at fixed
host mass, we find a substantial
reduction in the anisotropy of the satellites of blue MRS hosts compared to 
red MRS hosts due to the different methods by which the luminous host galaxies
have been embedded within their halos.

The locations of satellite galaxies with respect to the symmetry
axes of their hosts may, at first glance, seem to be a mere curiosity. 
However, the
current investigations are beginning to show that the locations of 
satellite galaxies can be used as direct probes of the large--scale
potentials of dark matter halos, and even provide clues to the
orientations of the host galaxies within their halos.    Out of necessity,
the resulting constraints are statistical in nature (since each
host galaxy generally has only 1 or 2 satellites), but this
makes the use of satellite galaxies as halo probes very complementary
to weak gravitational lensing techniques.  Considerably
larger samples of hosts and satellites than those used here may reveal a 
wealth of information about the sizes and shapes of the dark matter halos
of the hosts, the orientation of the hosts within their halos, and the
history of mass accretion by large, bright galaxies.

\section*{Acknowledgments}

It is a great pleasure to thank Simon White and the Max Planck Institute
for Astrophysics for hospitality and financial
support of a collaborative visit that allowed us to work directly
with the MRS particle files.   We are also very pleased to thank the
referee for thoughtful, constructive remarks that truly improved the manuscript.
Support under NSF contracts AST-0406844 and AST-0708468 is
gratefully acknowledged.  Funding for the SDSS has been provided by
the Alfred P. Sloan Foundation, the Participating Institutions, NASA,
the NSF, the US Department of Energy, the Japanese Monbukagakusho, and
the Max Planck Society.  The SDSS is managed by the Astrophysical
Research Consortium for the Participating Institutions (the University
of Chicago, Fermilab, the Institute for Advanced Study, the Japan 
Participation Group, The Johns Hopkins University, Los Alamos National
Laboratory, the Max Planck Institute for Astronomy, the Max Planck
Institute for Astrophysics, New Mexico State University, the
University of Pittsburgh, Princeton University, the US Naval Observatory,
and the University of Washington.  The SDSS Web site is 
\url{http://www.sdss.org}.

\clearpage

\centerline{Table 1: Numbers of Hosts and Satellites} 
\bigskip
\centerline{
\begin{tabular}{lcccc}
\hline\hline
\multicolumn{1}{c}{ } &
\multicolumn{2}{c}{SDSS} &
\multicolumn{2}{c}{MRS} \\
  & hosts & satellites & hosts & satellites \\ \hline
  primary sample (all galaxies) & 4,487  & 7,399  & 70,882 & 140,712 \\
galaxies with known $M_{\rm stellar}$ & 4,412 & 7,296 & 70,882 & 140,712 \\
galaxies with known SSFR & 2,421 & 4,004 & 47,157 & 79,812 \\
red galaxies & 2,926 & 2,334 & 37,022 & 86,178 \\
blue galaxies & 1,561 & 5,065 & 33,860 & 54,534 \\
\hline
\end{tabular}
}

\clearpage
\begin{figure}
\centerline{\scalebox{0.80}{\includegraphics{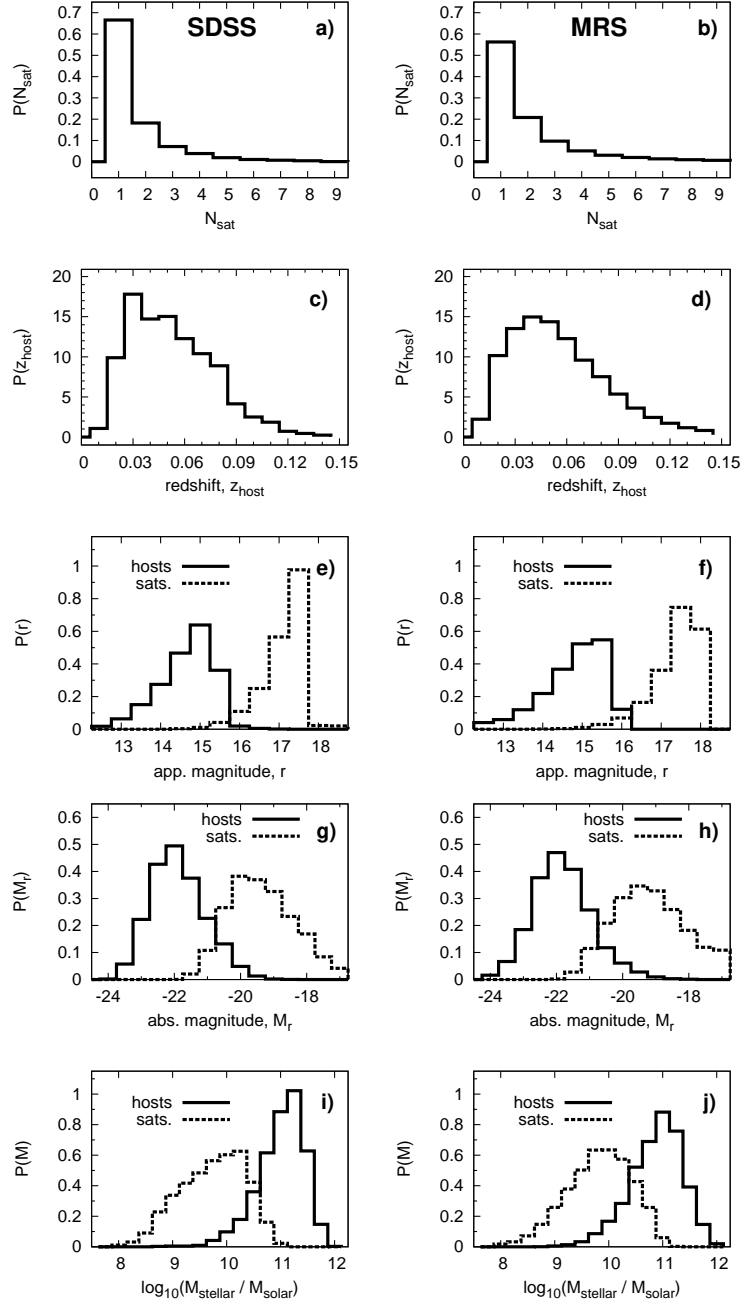} } }%
\vskip -0.0cm%
\caption{Summary of basic properties of the host--satellite
pairs in the SDSS (left panels) and the MRS (right
panels).  From top to bottom, the panels show probability distributions
for the number of satellites per host, the redshift distribution of the
hosts, the $r$--band apparent magnitude distributions of the hosts and satellites,
the $r$--band absolute magnitude distributions of the hosts and satellites,
and the distribution of stellar masses for the hosts and satellites.  In 
panels e) through j) dotted lines indicate results for the satellites and
solid lines indicate results for the hosts.}
\label{fig1}
\end{figure}

\begin{figure}
\centerline{\scalebox{0.90}{\includegraphics{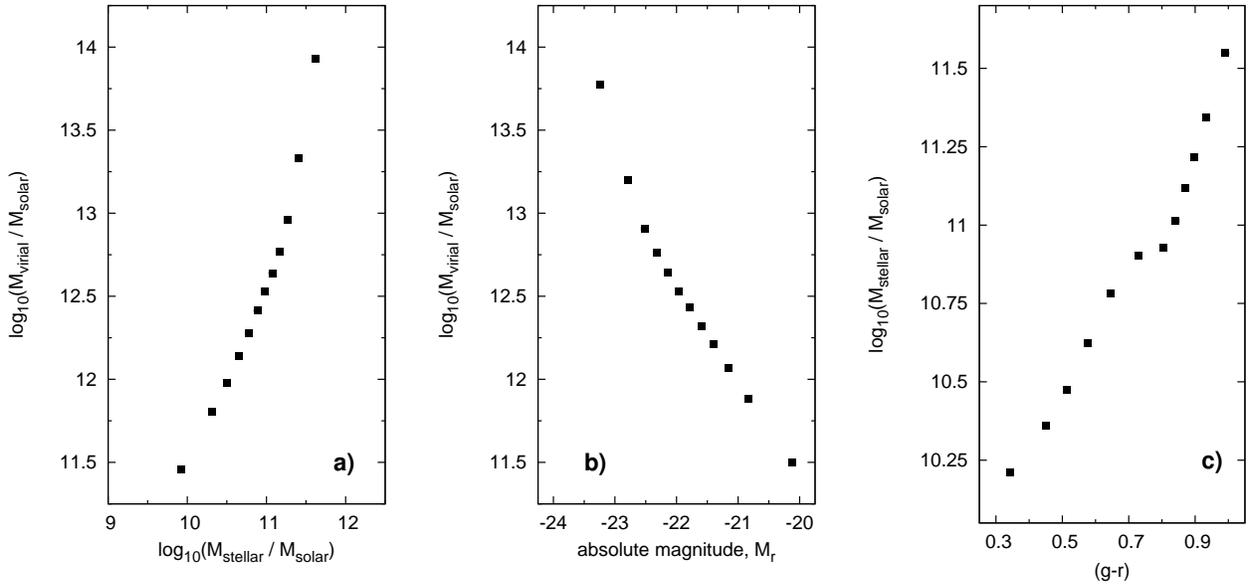} } }%
\vskip -0.0cm%
\caption{Properties of MRS host galaxies.  a) Mean host halo virial mass as a
function of stellar mass.  b) Mean host halo virial mass
as a function of absolute $r$--band magnitude.  c) Mean host stellar mass as a 
function of $(g-r)$, computed at $z=0$.  In all panels the data have been
binned such that there are an equal number of objects per bin.
In all cases the standard deviations in the mean values are comparable
to or smaller than the data points.
}
\label{fig2}
\end{figure}

\begin{figure}
\centerline{\scalebox{0.90}{\includegraphics{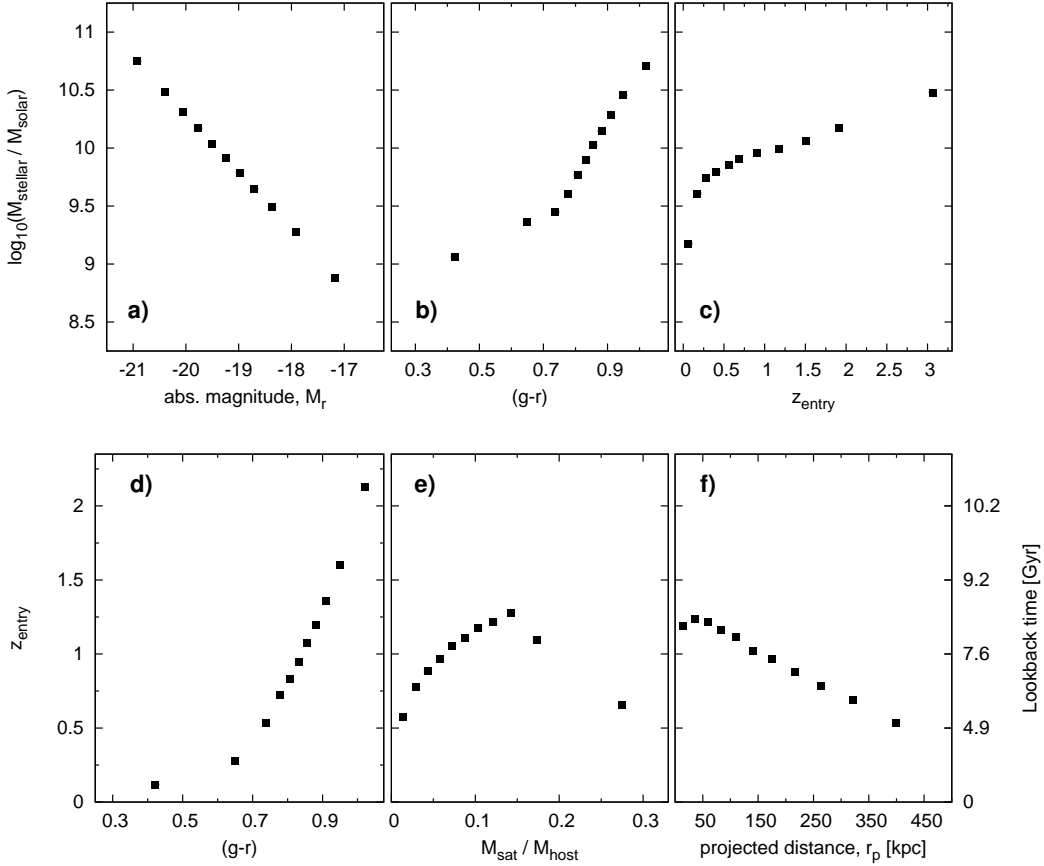} } }%
\vskip -0.0cm%
\caption{Properties of 
satellite galaxies in the MRS that are located within a physical distance
$r_{3D} \le 500$~kpc of a host galaxy. {\it Top:} Mean satellite stellar
mass as a function of absolute $r$--band magnitude (panel a), 
$(g-r)$ at $z=0$ (panel b), and
redshift at which the
satellite first entered its host's halo (panel c). 
{\it Bottom:} Mean redshift at
which a satellite first entered the halo of its host as a function of
$(g-r)$ at $z=0$ (panel d), ratio of satellite to host stellar mass
(panel e), and projected distance at which the satellite is found (panel f).
In each panel the data have been binned such that there are an equal number of objects
per bin.
In all cases the standard deviations in the mean values are comparable
to or smaller than the data points.}
\label{fig3}
\end{figure}

\begin{figure}
\centerline{\scalebox{1.00}{\includegraphics{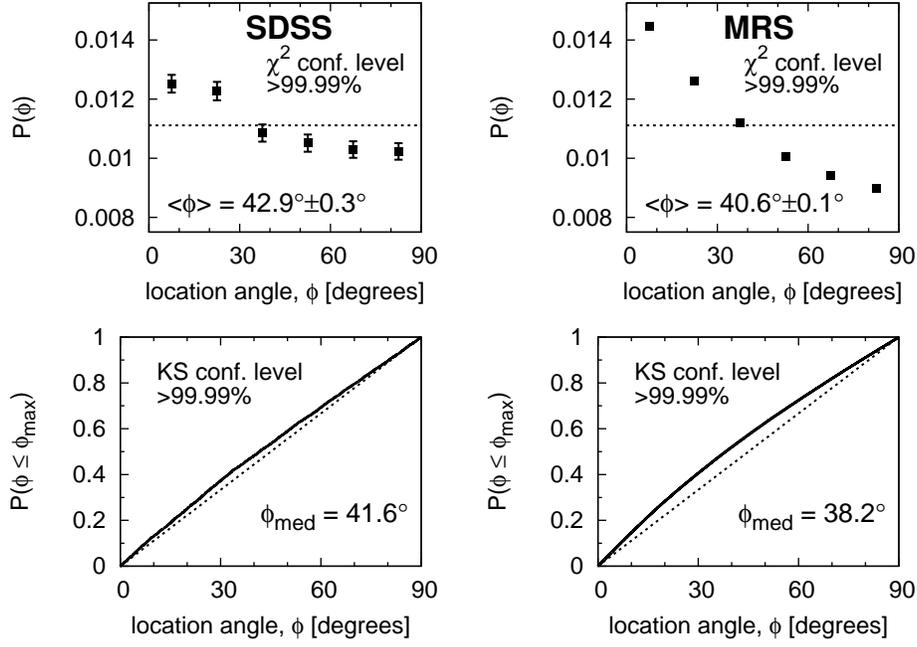} } }%
\vskip 0.0cm%
\caption{ {\it Top:} Differential probability distribution,
$P(\phi)$, for the locations of all satellites,  measured with
respect to the major axes of the hosts.
Dotted line shows the expectation for a uniform
(i.e., circularly--symmetric) distribution of satellites. 
The mean satellite location, $\left< \phi \right>$, and 
the confidence level at which the $\chi^2$ test rejects a uniform distribution
distribution are shown in the panels.
Error bars are omitted when they are
comparable to or smaller than the data point.
{\it Bottom:}
Cumulative probability distribution, $P(\phi \le \phi_{\rm max})$,
for the locations
of the satellites with respect to the major
axes of the hosts (solid line).  Also shown is
$P(\phi \le \phi_{\rm max})$ for a uniform distribution (dotted line).
The median satellite location, $\phi_{\rm med}$, and 
the confidence level at which the KS test rejects a uniform distribution 
are shown in the panels.
{\it Left:} Satellites in the SDSS.  {\it Right:} Satellites in the
MRS.
All satellites with $r_p \le 500$~kpc have been used in the calculations.}
\label{fig4}
\end{figure}

\begin{figure}
\centerline{\scalebox{1.00}{\includegraphics{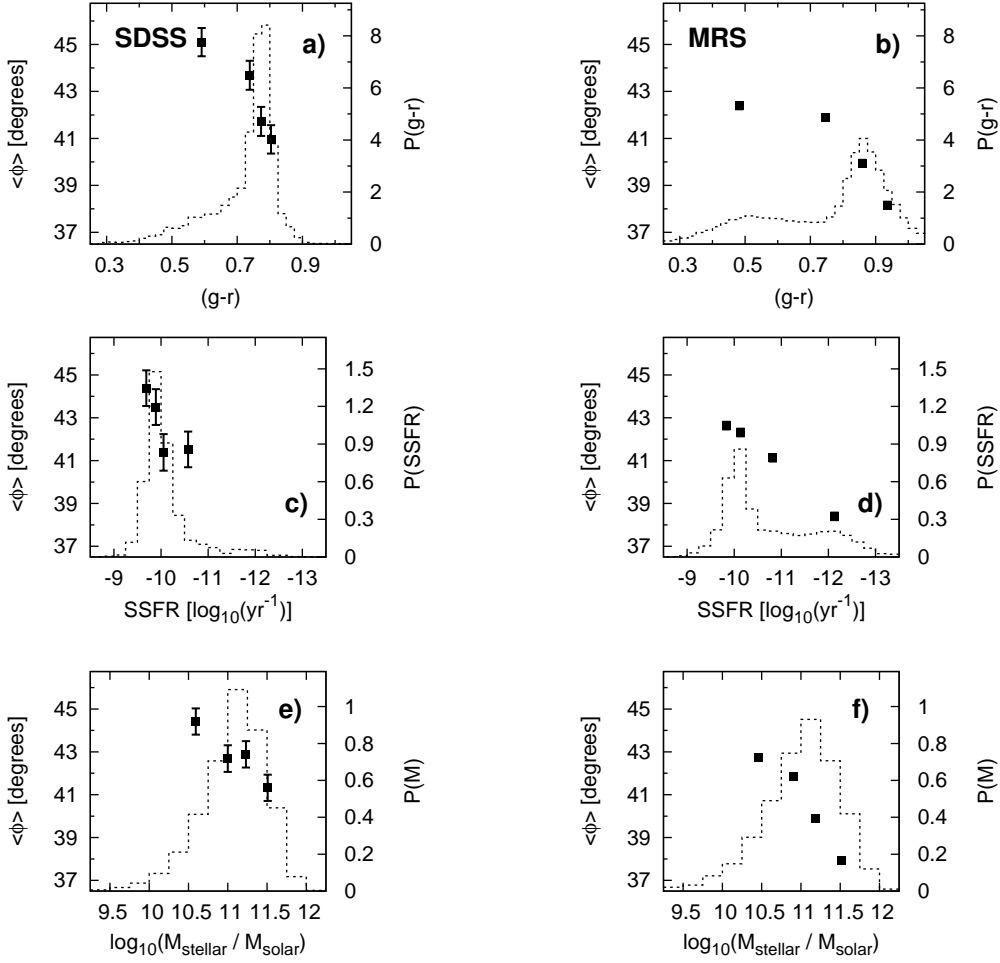} } }%
\vskip 0.0cm%
\caption{Data points with error bars show the
mean satellite location, $\left< \phi \right>$,
for SDSS satellites
(left panels) and MRS satellites (right panels), as a function of
various properties of the hosts.  
Histograms show the distribution
of the host property in each panel.
{\it Top:} $\left< \phi \right>$ as a function of the host's
$(g-r)$ color, computed at $z=0$.  {\it Middle:} $\left< \phi \right>$ as a function
of host
specific star formation rate, SSFR.  {\it Bottom:} $\left< \phi \right>$
as function of host stellar mass. 
All satellites with $r_p \le 500$~kpc have been used in the calculations.
In each panel
the data have been binned such that there are an equal number of 
objects per bin in the calculation of $\left< \phi \right>$.  
Error bars are omitted when the standard
deviation in the mean value of $\phi$ is smaller than
the data point.}
\label{fig5}
\end{figure}

\begin{figure}
\centerline{ \scalebox{1.00}{\includegraphics{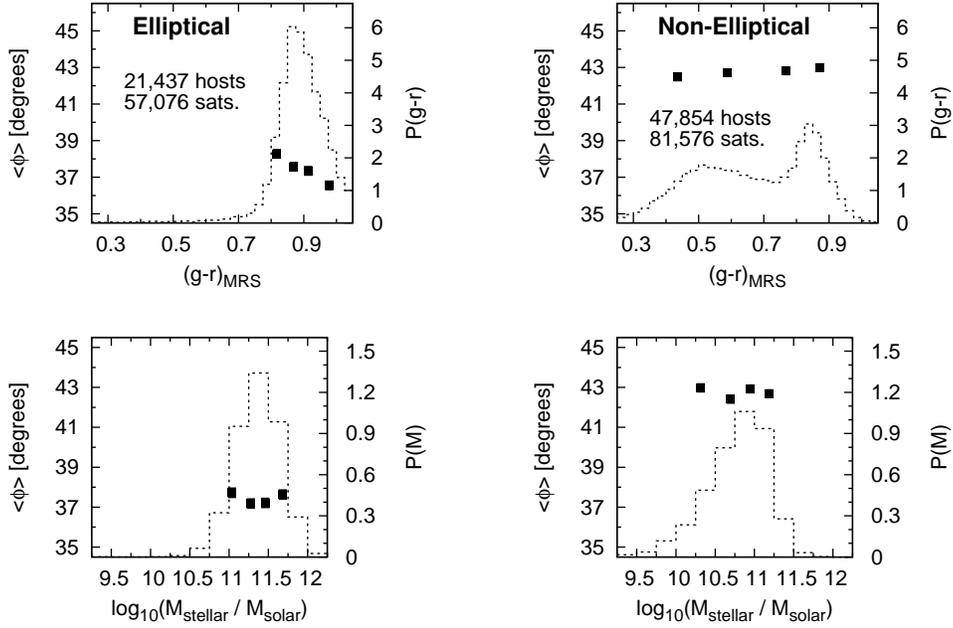} } }%
\vskip 0.0cm%
\caption{Data points show the mean satellite location, 
$\left< \phi \right>$, for MRS satellites
as a function of host properties for elliptical MRS hosts (left
panels) and non--elliptical MRS hosts (right panels).  
Histograms show the distribution
of the host property in each panel.  {\it Top:} Mean satellite location 
as a function of host $(g-r)$ color.  {\it Bottom:} Mean 
satellite location as a function of host stellar mass.
All satellites with $r_p \le 500$~kpc have been used in the calculations.
In all panels the data have been binned such that there are an equal
number of objects per data point.
In all cases the standard deviation in the mean value of $\phi$ is
comparable to or smaller than the data points.
}
\label{fig6}
\end{figure}

\begin{figure}
\centerline{\scalebox{1.00}{\includegraphics{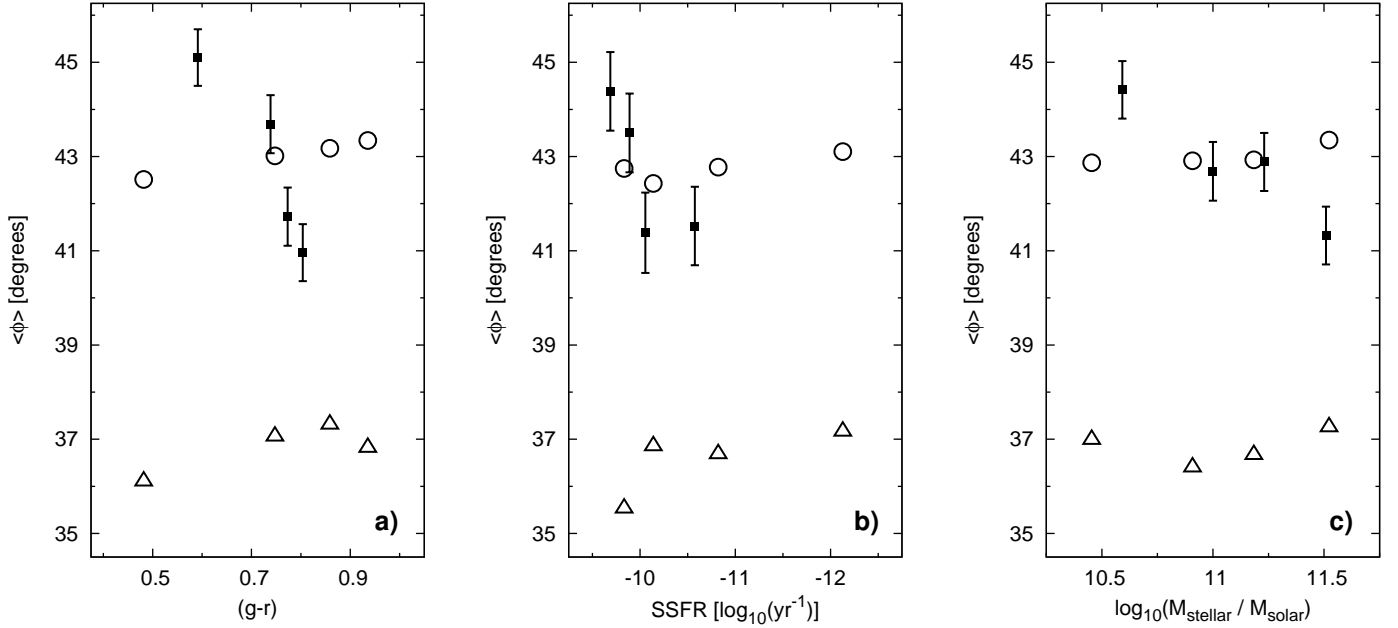} } }%
\vskip 0.0cm%
\caption{Same as Fig.~5, except here single image 
assignment prescriptions are used to define the major axes of the MRS hosts.
Open circles: Major axes of all MRS hosts are obtained from
projections of circular disks onto the sky, where the angular
momenta of the disks are aligned with the angular momenta of the
halos.  
Open triangles: Major axes of all MRS hosts are 
obtained from projections of the halo equivalent ellipsoids onto
the sky.  
Solid squares: SDSS results from Fig.~5. 
Error bars are omitted when the standard deviation in the
mean value of $\phi$ is comparable to or smaller than the data point.
}
\label{fig7}

\end{figure}
\begin{figure}
\centerline{\scalebox{1.00}{\includegraphics{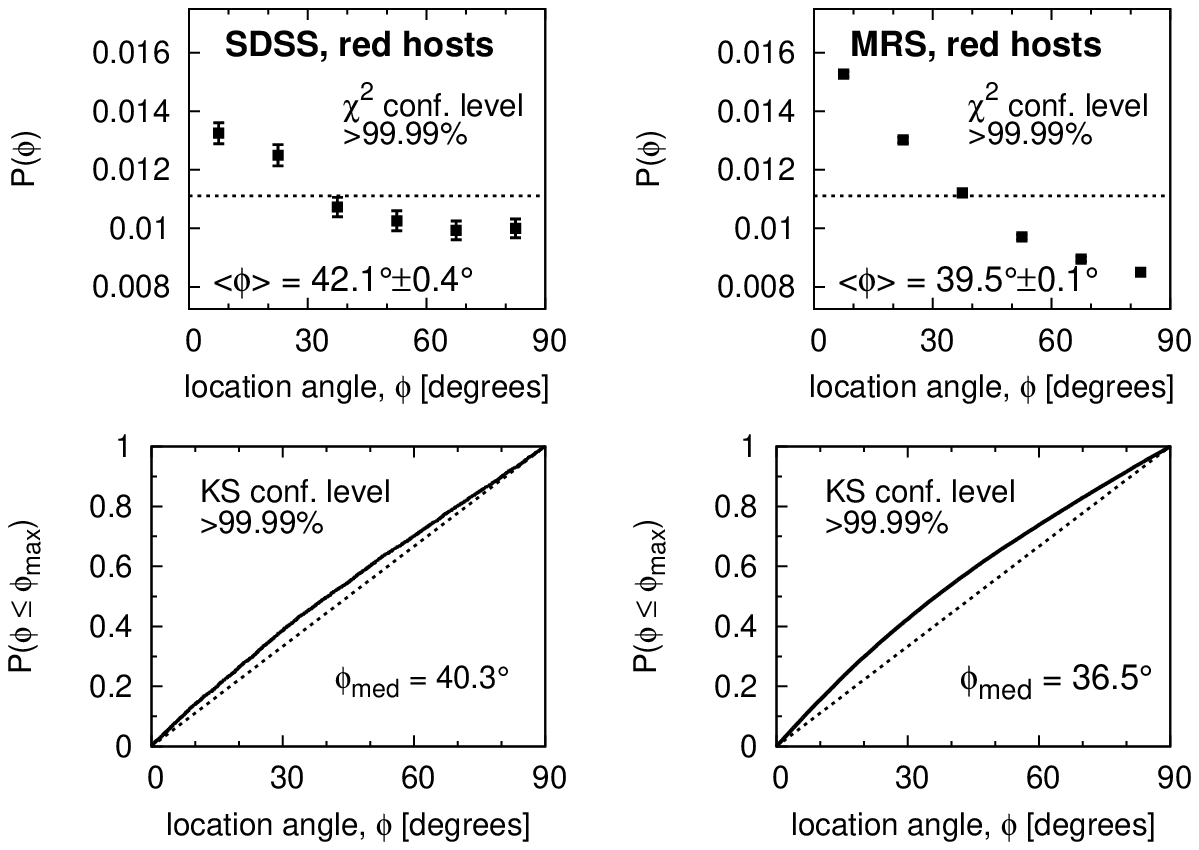} } }%
\vskip 0.0cm%
\caption{Same as Fig.~4, but for the satellites of red
hosts. 
All satellites with projected distances $r_p \le 500$~kpc
have been used in the calculations.
}
\label{fig8}
\end{figure}

\begin{figure}
\centerline{\scalebox{1.00}{\includegraphics{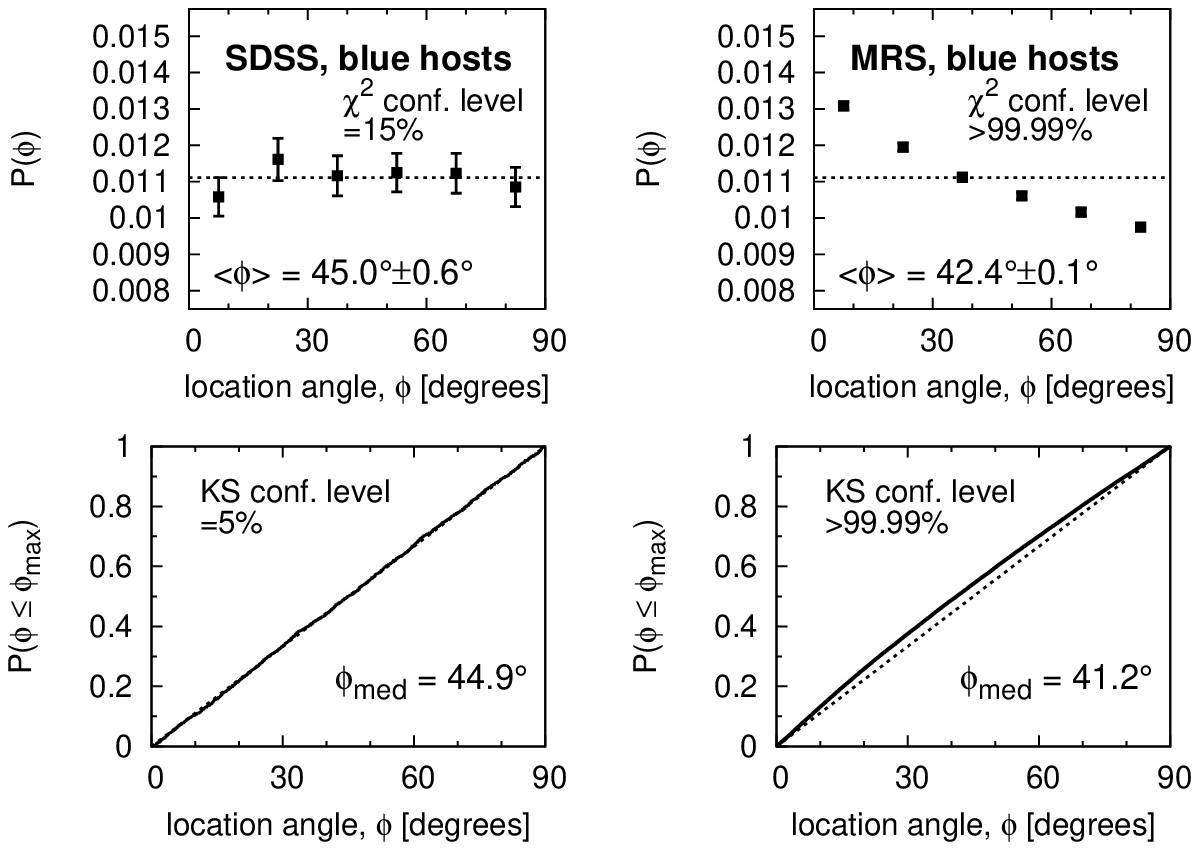} } }%
\vskip 0.0cm%
\caption{ Same as Fig.~4, but for the satellites of
blue hosts.  
All satellites with projected distances $r_p \le
500$~kpc have been used in the calculation.
}
\label{fig9}
\end{figure}

\begin{figure}
\centerline{\scalebox{0.90}{\includegraphics{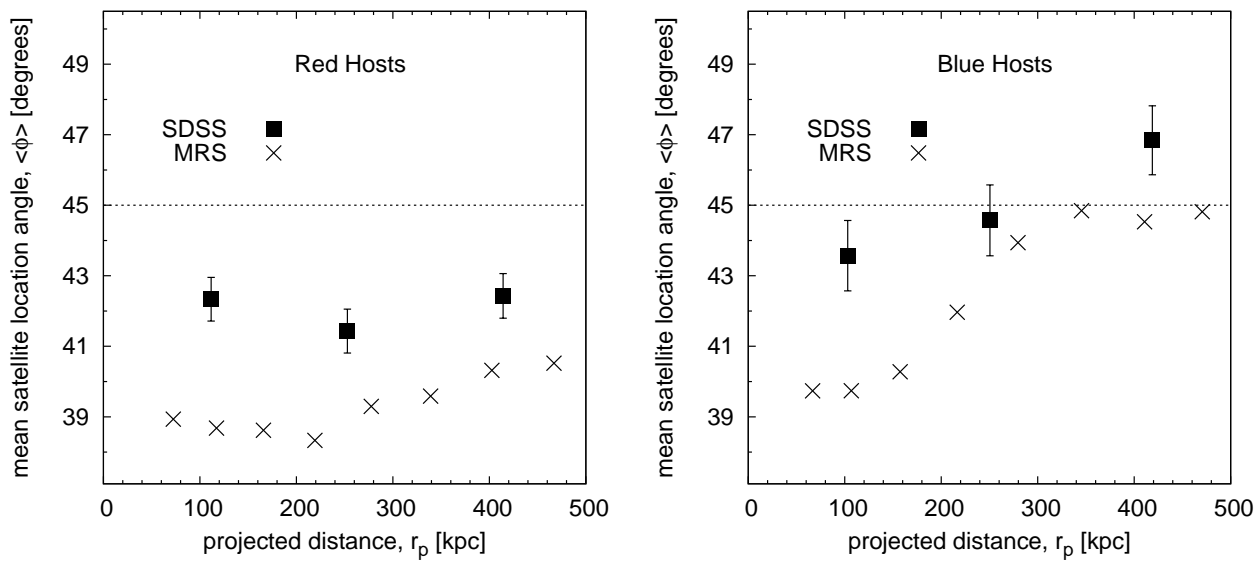} } }%
\vskip 0.0cm%
\caption{Mean satellite location as a function of projected
distance, $r_p$, for the satellites of SDSS hosts (solid squares) and
MRS hosts (crosses).  {\it Left: } Satellites of red hosts.
{\it Right:} Satellites of blue hosts.  Error bars are omitted when
the standard deviation in the mean value of $\phi$ is comparable
to or smaller than
the data point.
}
\label{fig10}
\end{figure}

\begin{figure}
\centerline{ \scalebox{1.00}{\includegraphics{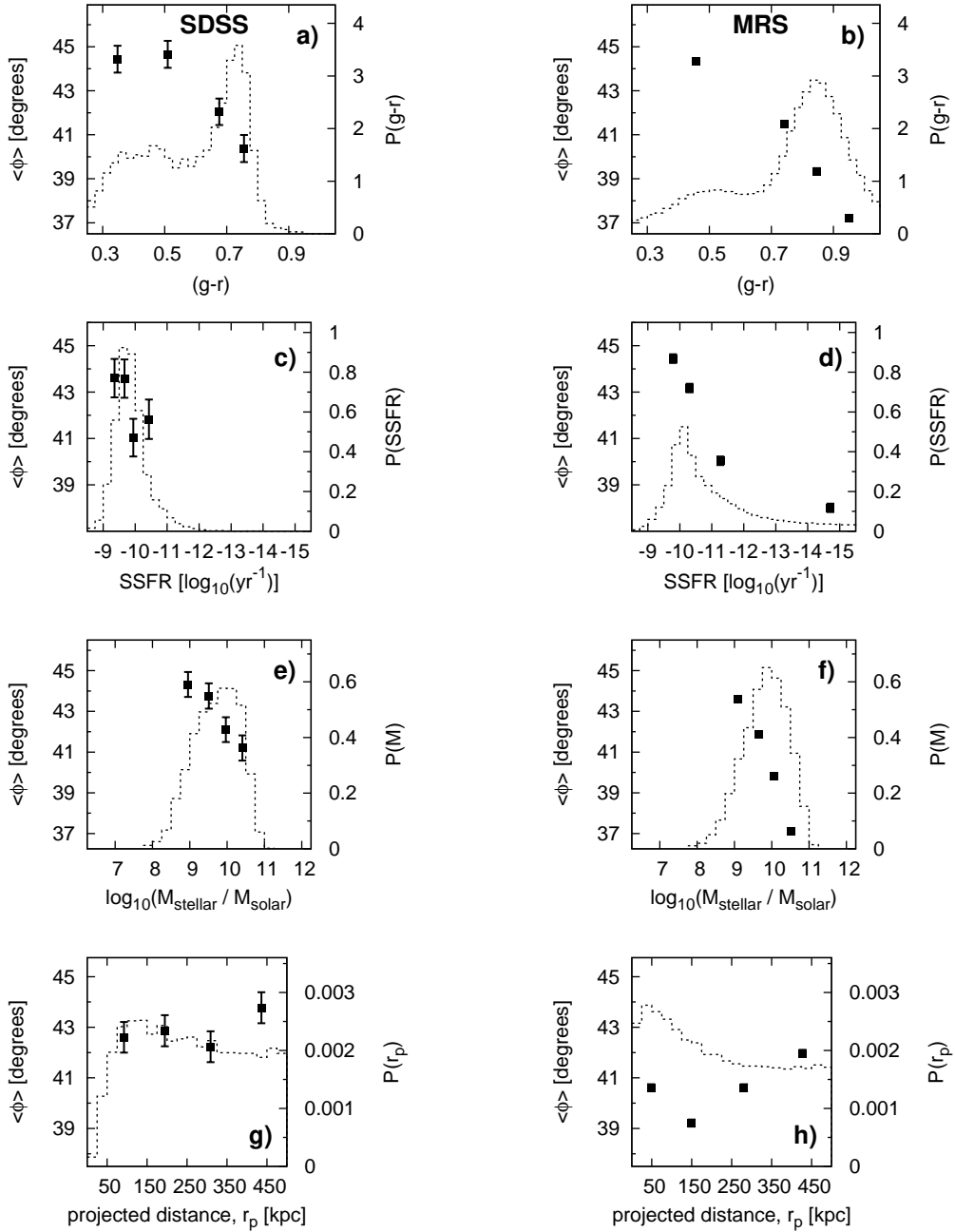} } }%
\vskip -0.5cm%
\caption{Data points with error bars show the
 mean satellite location, $\left< \phi \right>$, for SDSS satellites
(left panels) and MRS satellites (right panels), as a function of
various properties of the satellites.  
Histograms show the distribution of the satellite property in each 
panel.  From top to bottom the panels show
 $\left< \phi \right>$ as a function of $(g-r)$,
$\left< \phi \right>$ as a function of satellite
specific star formation rate (SSFR), $\left< \phi \right>$ as
as a function of satellite stellar mass, and 
$\left< \phi \right>$ as a function of the projected
distance at which the satellites are found. 
In each panel the data have been binned such that there are an equal 
number of objects per bin in the calculation of $\left< \phi \right>$.  
Error bars are omitted when the standard
deviation in the mean value of $\phi$ is comparable to or
smaller than the data point.}
\label{fig11}
\end{figure}

\begin{figure}
\centerline{ \scalebox{0.90}{\includegraphics{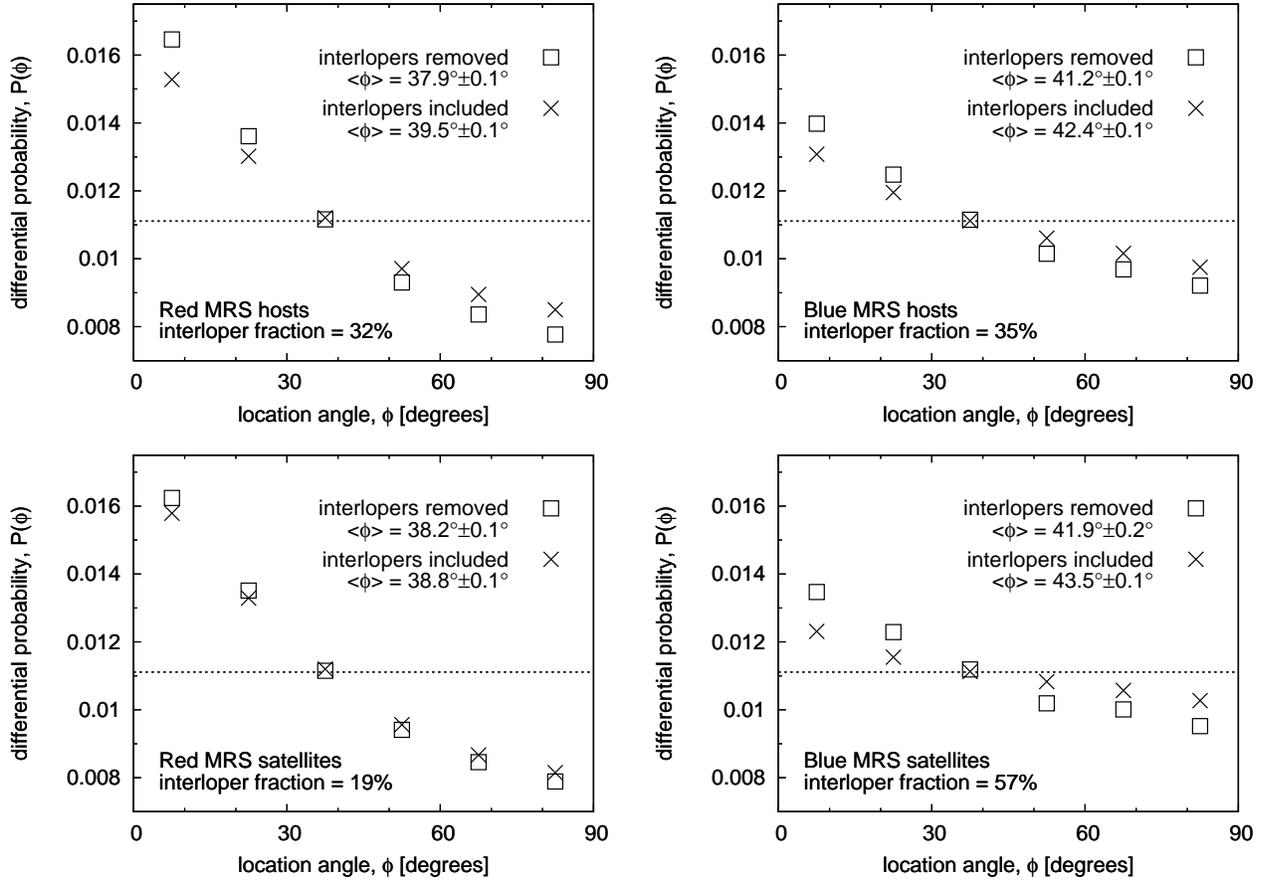} } }%
\caption{
Effects of interlopers on the satellite locations in the MRS.   Open
points show $P(\phi)$ using all objects that were identified as satellites
according to the selection criteria in \S3.1.  In all cases the
error in $P(\phi)$ is smaller than the data points. Solid points show 
$P(\phi)$ after all interlopers have been removed from the 
satellite sample (see text).  {\it Top panels:} 
$P(\phi)$ for red (left) and blue (right) MRS hosts.  {\it Bottom panels:}
$P(\phi)$ for red (left) and blue(right) MRS satellites.
}
\label{fig12}
\end{figure}

\begin{figure}
\centerline{ \scalebox{1.00}{\includegraphics{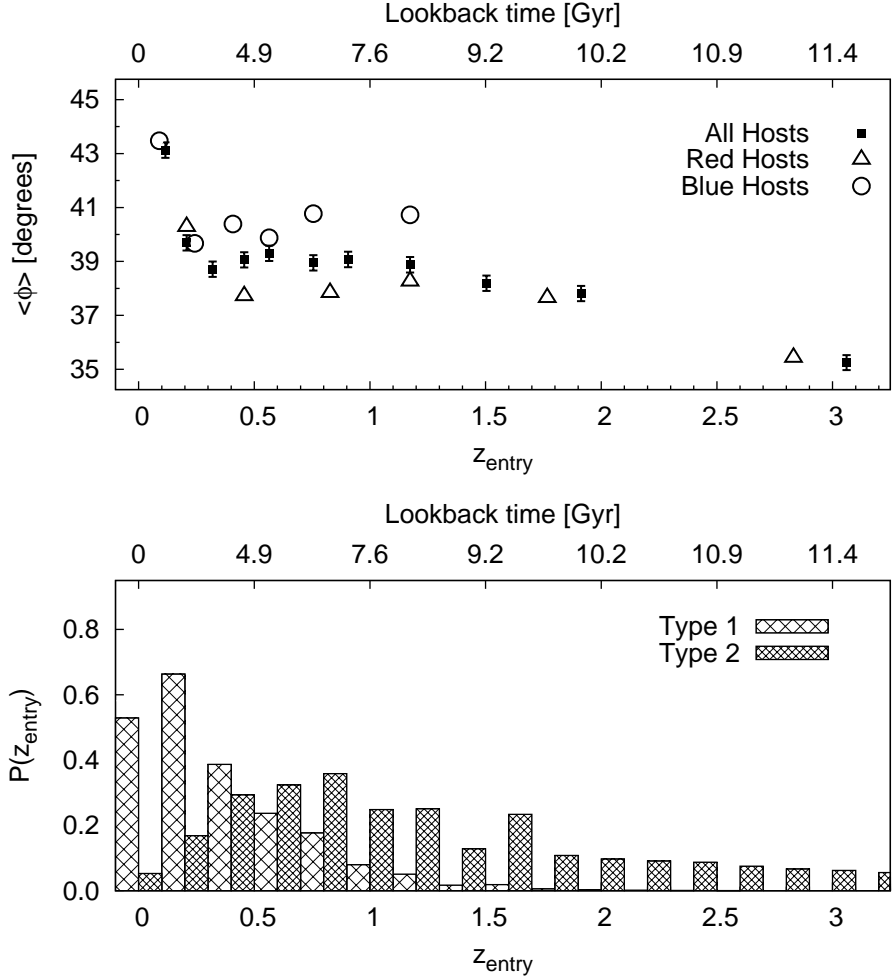} } }%
\vskip 0.0cm%
\caption{{\it Top:} Mean satellite location at $z=0$ for genuine MRS satellites as a 
function of the redshift at which they first entered their host's halo.  
Here all satellites are located within a physical distance
of $r_{3D} \le 500$~kpc of the host at the
present day. The data have been binned such that there are an 
equal number of objects per bin, and error bars are omitted when the
standard deviation in the mean value of $\phi$ is comparable to or smaller than
the data point. Solid squares: satellites of all MRS hosts.  Open circles:
satellites of blue MRS hosts.  Open triangles: satellites of red MRS hosts.
{\it Bottom:} Probability distribution for the redshift
at which the genuine MRS satellites first entered their host's halo. 
}
\label{fig13}
\end{figure}

\end{document}